\documentclass[11pt]{article}

\usepackage{amsmath,amsfonts,amssymb,amsthm}
\usepackage{authblk}
\usepackage{fullpage}
\usepackage{times}
\usepackage{hyperref}
\usepackage{microtype}
\usepackage{color}
\usepackage{cleveref}
\usepackage[ruled,vlined]{algorithm2e}

\crefformat{footnote}{#2\footnotemark[#1]#3}

\definecolor{light-gray}{gray}{0.5}

\crefformat{footnote}{#2\footnotemark[#1]#3}

\definecolor{light-gray}{gray}{0.5}

\newcommand{\rimplies}{\ensuremath{\Rightarrow_R}}

\newcommand{\B}{\ensuremath{\mathbb{B}}}

\newcommand{\N}{\ensuremath{\mathbb{N}}}
\newcommand{\Q}{\ensuremath{\mathbb{Q}}}
\newcommand{\R}{\ensuremath{\mathbb{R}}}

\newcommand{\Tr}{\ensuremath{\mathsf{T}}}
\newcommand{\Z}{\ensuremath{\mathbb{Z}}}

\numberwithin{equation}{section}

\DeclareMathOperator{\poly}{poly}

\newcommand{\eps}{\varepsilon}

\newcommand{\va}{\mathbf{a}}
\newcommand{\vb}{\mathbf{b}}
\newcommand{\vc}{\mathbf{c}}
\newcommand{\vd}{\mathbf{d}}
\newcommand{\ve}{\mathbf{e}}

\newcommand{\vv}{\mathbf{v}}

\newcommand{\vx}{\mathbf{x}}
\newcommand{\vy}{\mathbf{y}}
\newcommand{\vz}{\mathbf{z}}
\newcommand{\vLa}{\boldsymbol\lambda}
\newcommand{\vEps}{\boldsymbol\varepsilon}

\newcommand{\CC}{\mathcal{C}}

\newcommand{\CL}{\mathcal{L}}

\newcommand{\CT}{\mathcal{T}}
\newcommand{\MA}{\mathsf{A}}

\newcommand{\MC}{\mathsf{C}}

\newcommand{\CG}{\mathrm{CG}}

\newcommand{\inner}[1]{\langle{#1}\rangle}

\newcommand{\set}[1]{\{{#1}\}}

\newcommand{\round}[1]{\lfloor{#1}\rceil}
\newcommand{\floor}[1]{\lfloor{#1}\rfloor}
\newcommand{\ceil}[1]{\lceil{#1}\rceil}
\newcommand{\len}[1]{\lvert{#1}\rvert}
\newcommand{\lenfit}[1]{\left\lvert{#1}\right\rvert}
\newcommand{\length}[1]{\lVert{#1}\rVert}
\newcommand{\lengthfit}[1]{\left\lVert{#1}\right\rVert}

\theoremstyle{plain}            
\newtheorem{theorem}{Theorem}[section]
\newtheorem{lemma}[theorem]{Lemma}
\newtheorem{corollary}[theorem]{Corollary}

\newtheorem{claim}[theorem]{Claim}

\theoremstyle{definition}       
\newtheorem{definition}[theorem]{Definition}

\theoremstyle{remark}           
\newtheorem{remark}[theorem]{Remark}

\begin{document}

\title{On the Complexity of Branching Proofs}
\author[1]{Daniel Dadush\thanks{Supported by ERC Starting Grant QIP--805241.} }
\author[1]{Samarth Tiwari}
\affil[1]{Centrum Wiskunde \& Informatica, Amsterdam 
\protect\\ {\tt \{dadush,samarth.tiwari\}@cwi.nl}}

\maketitle

\begin{abstract}
We consider the task of proving integer infeasibility of a bounded convex $K$ in
$\R^n$ using a general branching proof system. In a general branching proof, one
constructs a branching tree by adding an integer disjunction $\va \vx \leq b$ or
$\va \vx \geq b+1$, $\va \in \Z^n$, $b \in \Z$, at each node, such that the leaves
of the tree correspond to empty sets (i.e., $K$ together with the inequalities
picked up from the root to leaf is empty). 

Recently, Beame et al (ITCS 2018), asked whether the bit size of the
coefficients in a branching proof, which they named {\rm stabbing planes} (SP)
refutations, for the case of polytopes derived from SAT formulas, can be assumed
to be polynomial in $n$. We resolve this question in the affirmative, by showing
that any branching proof can be recompiled so that the normals of the
disjunctions have coefficients of size at most $(n R)^{O(n^2)}$, where $R \in \N$ is
the radius of an $\ell_1$ ball containing $K$, while increasing the number
of nodes in the branching tree by at most a factor $O(n)$. Our recompilation
techniques works by first replacing each disjunction using an iterated
Diophantine approximation, introduced by Frank and Tardos (Combinatorica 1986),
and proceeds by ``fixing up'' the leaves of the tree using judiciously added
Chv{\'a}tal-Gomory (CG) cuts. 

As our second contribution, we show that Tseitin formulas, an important class of
infeasible SAT instances, have quasi-polynomial sized cutting plane (CP)
refutations. This disproves a conjecture that Tseitin formulas are
(exponentially) hard for CP. Our upper bound follows by recompiling the
quasi-polynomial sized SP refutations for Tseitin formulas due to Beame et al,
which have a special enumerative form, into a CP proof of at most twice the
length using a serialization technique of Cook et al (Discrete Appl. Math.
1987). 

As our final contribution, we give a simple family of polytopes in $[0,1]^n$
requiring branching proofs of length $2^n/n$.
\end{abstract}


\section{Introduction}

A principal challenge in SAT solving is finding short proofs of unsatisfiability
of SAT formulas. This task is particularly important in the automatic
verification of computer programs, where incorrect runs of the program or bugs
(e.g., divide by zero) can be encoded as satisfying assignments to SAT formulas
derived from the program specification. In this case, the corresponding formula
is an UNSAT instance if the corresponding program is correct, or at least devoid
of certain types of bugs. 

The study of how long or short such UNSAT proofs can be is the main focus of the
field of proof complexity. Indeed, popular SAT algorithms, such as DPLL search,
i.e.~branching on variables combined with unit propagation, or Conflict Driven
Clause Learning (CDCL), implicitly generate infeasibility proofs in standard
proof systems such as Resolution or Cutting Planes. From the negative
perspective, lower bounds on the length of UNSAT proofs in these systems
automatically imply lower bounds on the running time of the corresponding SAT
algorithms. On the positive side, understanding which UNSAT instances have short
proofs can inspire the design of good heuristics and algorithms for trying to
find such proofs automatically.   

The analogous problem in the context of Integer Programming (IP) is that of
showing that a linear system of inequalities has no integer solutions. This
problem also encapsulates SAT: for a formula $\Phi(\vx) := \wedge_{j \in [m]}
C_j(\vx)$, where $C_j(\vx) = \vee_{i \in L_j} x_i \vee_{i \in \bar{L}_j} \bar{x}_i$,
$j \in [m]$, $\Phi$ is unsatisfiable if and only if the linear system
\begin{align}
\tag{SAT-LP}
\sum_{i \in L_j} x_i + \sum_{i \in \bar{L}_j} (1-x_i) &\geq 1, j \in [m]
\label{SAT-LP} \\ 
0 \leq x_i &\leq 1, i \in [n] \nonumber
\end{align}

has no integer solutions (in this case $\set{0,1}$). IP solvers such as CPLEX or
Gurobi routinely produce such infeasibility proofs in the so-called proof of
optimality phase of the solution process. More precisely, once a solver has
found a candidate optimal solution $\vx^*$ to an integer linear program 
\begin{equation}
\tag{IP} \label{IP}
\min \vc \vx \quad \text{subject to } \MA \vx \leq \vb, \vx \in \Z^n
\end{equation}
optimality is proved by showing that the linear system
\begin{align}
\tag{IP-LP}
\vc \vx &< \vc \vx^* \label{IP-LP} \\
\MA \vx &\leq \vb \nonumber
\end{align}
has no integer solutions. In practice, this is most often achieved by a mixture
of Branch \& Bound and Cutting Planes. We note that most applications are
modeled using \emph{mixed} integer linear programs (MIP), where a decision
variable $x_i$ can be continuous ($x_i \in \R$), binary ($x_i \in \{0,1\}$) or
general integer ($x_i \in \Z$), with binary and continuous variables being the
most common.

\subsection{Branching Proofs}
\label{subsection:branching-proofs}
For proving infeasibility of a SAT formula or an integer linear program, where
we denote the continuous relaxation of the feasible region by $K \subseteq \R^n$
(e.g.,~\eqref{SAT-LP} or~\eqref{IP-LP}), the most basic strategy is to build a
search tree based on so-called variable branching. That is, we build a rooted
binary tree $\CT$, where at each internal node $v$ we choose a ``promising''
candidate integer variable $x_i$ and create two children $v_l, v_r$
corresponding either side of the disjunction $x_i \leq b$ (left child $v_l$) and
$x_i \geq b+1$ (right child $v_r$), for some $b \in \Z$.  The edge from the
parent to its child is labeled with the corresponding inequality. If $x_i$ is
binary, one always sets $b=0$, corresponding to branching on $x_i = 0$ or $x_i =
1$. To each node is associated its continuous relaxation $K_v$, corresponding to
$K$ together with the inequalities on the edges of the unique path from the root
to $v$ in $\CT$. To be a valid proof of integer infeasibility, we require that
the continuous relaxation $K_v$ be empty at every leaf node $v \in \CT$. We then
call the proof tree $\CT$ as above a \emph{variable} branching proof of integer
infeasibility for $K$. We will consider the length of branching proof,
interpreted as the ``number of lines'' of the proof, to be equal to the number
of nodes in $\CT$, which we denote $|\CT|$. 

When applied to a SAT formula as in~\eqref{SAT-LP}, a variable branching tree
$\CT$ as above is in correspondence with a run of DPLL search, noting that LP
infeasibility of a node is equivalent to unit propagation (i.e.,~iteratively
propagating the values of variables appearing in single literal clauses)
yielding a conflict\footnote{Note that if unit propagation finds a conflict at
a node of the tree, the corresponding node LP (i.e.~\eqref{SAT-LP} with some
variables fixed to $0$ or $1$) is also
infeasible. If unit propagation terminates without a conflict, then setting all
non-propagated variables to $1/2$ yields a feasible LP solution since every
surviving clause has at least $2$ literals.}. Similarly when
applied to an integer program as in~\eqref{IP-LP} for which the optimal value is
known, the above is equivalent to standard Branch and Bound. 

\subparagraph*{Branching on General Integer Disjunctions} To obtain a more general
proof strategy one may examine a richer class of disjunctions. Instead of
branching only on variables as above, one may also branch on a general
integer disjunction $\va \vx \leq b$ or $\va \vx \geq b+1$, where $\va \in \Z^n$
and $b \in \Z$, noting that any integer point $\vx \in \Z^n$ must satisfy
exactly one of these inequalities. One may then define branching proofs of
infeasibility for $K$ using general integer disjunctions exactly as above, which
we call \emph{general} branching proofs. We note that in principle, the
continuous relaxation $K$ can be arbitrary, i.e.~it need not be a
polytope. In this work, we will in fact consider the case where $K$ is a compact
convex set in $\R^n$. Furthermore, it is easy to extend branching proofs to the
case of \emph{mixed} integer infeasibility, where we want to certify that $K
\cap \Z^k \times \R^{n-k} = \emptyset$, that is, where only the first $k$
variables are restricted to be integer. In this setting, one need only restrict
the disjunctions $\va \vx \leq b$ or $\geq b$ to have support on the integer
variables; precisely, we enforce $\va \in \Z^k \times \{0\}^{n-k}, b \in \Z$.

As formalized above, the attentive reader may have noticed that there is no
mechanism to ``certify'' the emptiness of the leaf nodes of the tree. In many
cases, such certificates can be appended to the leaves yielding a
\emph{certified} branching proof, however their exact form will differ depending
on the representation of $K$ (e.g., LP, SOCP or SDP). In the important
case where the continuous relaxation is a polytope $K = \set{\vx \in \R^n: \MC \vx
\leq \vd}$, emptiness of a leaf node can indeed be certified efficiently using a
so-called Farkas certificate of infeasibility. Let $\CT$ be branching proof for
$K$ and let $v \in \CT$ be a leaf node with $K_v = \set{\vx \in \R^n: \MC \vx
\leq \vd, \MA_v \vx \leq \vb_v}$, where $\MA_v \vx \leq \vb_v$ represents all
the inequalities induced by the branching decisions on the path from the root to
$v$. Then, by Farkas's lemma $K_v = \emptyset$ iff there exists multipliers
$\vLa_v := (\vLa_{v,1},\vLa_{v,2}) \geq 0$, known as a \emph{Farkas
certificate}, such that $\vLa_{v,1} \MC + \vLa_{v,2} \MA_{v} = 0$ and $\vLa_{v,1}
\vd + \vLa_{v,2}\vb < 0$. Therefore, for a polyhedral feasible region, we may
certify the branching proof by labeling each leaf node $v \in \CT$ with its
Farkas certificate $\vLa_v$.

For a variable branching proof $\CT$, especially for $\set{0,1}$ IPs, the tree
size $|\CT|$ is arguably the most important measure of the complexity of the
proof. However, for a general branching proof $\CT$, the tree size $|\CT|$
ignores the ``complexity'' of the individual disjunctions. Note that we have not
a priori set any restrictions on the size of the coefficients for the
disjunctions $\va \vx \leq b$ or $\geq b+1$ used in the nodes of the tree. To
accurately capture this complexity, we will also measure the number of bits
needed to write down the description of $\CT$, which we denote by $\inner{\CT
}$. Here, $\inner{\CT}$ includes the bit-size of all the disjunctions $\va \vx
\leq b$ or $\geq b+1$ used in $\CT$. For a certified branching proof, as
introduced above, we also include the bit-length of the infeasibility
certificates at the leaves to $\langle \CT \rangle$. Understanding how large the
coefficients need to be to ensure near-optimal tree size will be one of the
principal interests of this work.   


\subparagraph*{Applications of General Branching} While variable branching is
the most prevalent in practice, due to its simplicity and ease of
implementation, it is well-known that branching on general integer disjunctions
can lead to much smaller search trees. In practice, general branching is used
when certain simple constraints such as $\sum_{i=1}^n x_i = 1$, $x_i$ binary,
are present in the model, which is part of the family of specially ordered set
constraints~\cite{BT70}. In this context, one may branch on $\sum_{i=1}^{n/2}
x_i = 0$ or $\sum_{i=1}^{n/2} x_i = 1$ to a get a more balanced search tree. A
more recent idea of Fischetti and Lodi~\cite{FL03}, known as local branching, is
to branch on disjunctions which control the Hamming distance to the best
incumbent solution $\vx^*$, e.g.~$\sum_{i: x_i^*=0} x_i + \sum_{i: x_i^*=1}
(1-x_i) \leq k$ or $\geq k+1$. This provides a very effective way of controlling
the search neighborhood, and allows one to find improving solutions more
quickly.    

From the theoretical side, a seminal result is that of Lenstra~\cite{Lenstra83},
who gave a fixed dimension polynomial time algorithm for Integer Programming
based on basis reduction and general branching. Relating to branching proofs,
his result directly implies that every integer free compact convex set admits a
general branching proof of length $O(f(n)^n)$, where $f(n)$, the so-called
flatness constant, is the supremum of the lattice width over integer free
compact convex set in dimension $n$. It is known that $f(n) =
\tilde{O}(n^{4/3})$~\cite{Bana96,Rudelson00} and $f(n) = \Omega(n)$. We note
that already in $\R^2$, there are simple integer free polytopes, e.g., $\set{\vx
\in \R^2: x_1-x_2 = 1/2, 0 \leq x_1 \leq k}$, for $k \in \N$, with arbitrarily
long variable branching proofs. Inspired by Lenstra's result, there has been a
line of work on the use of basis reduction techniques to reformulate IPs so that
they become ``easy'' for variable branching. This approach has been successfully
theoretically analyzed for certain classes of knapsack problems as well as
random IPs (see~\cite{PT10} for a survey) and experimentally analyzed on various
classes of instances~\cite{AL04,KP09}. There has also been experimental work on
how to come up with good general branching directions in practice using
heuristic methods~\cite{OM01,MR09,KC11}.

\subsection{Cutting Planes} 

Another fundamental proof system, studied extensively within both the IP and SAT
contexts are cutting planes (CP) proofs.  The most fundamental class of cutting
planes are so-called Chv{\'a}tal-Gomory (CG) cuts, which are the principal class
studied within SAT and one of the most important classes of cuts in
IP~\cite{jour/bams/Gomory58}. 

CG cuts for a set $K \subseteq \R^n$ are derived geometrically as follows.
Assume that the inequality $\va \vx \leq r$, $\va \in \Z^n$, $r \in \R$, is
valid for $K$, that is, $\vx \in K \Rightarrow \va \vx \leq r$. Then, the
inequality $\va \vx \leq \floor{r}$ is valid for $K \cap \Z^n$, since $\vx \in
\Z^n$ implies that $\va \vx \in \Z$. Given $\va \in \Z^n$, the
strongest cut of this form one can derive for $K$ is clearly $\va \vx \leq
\floor{\sup_{\vz \in K} \va \vz}$. We therefore denote this cut to be the CG cut
of $K$ induced by $\va$, and we use the notation $\CG(K,\va) := \set{\vx \in K:
\va \vx \leq \floor{\sup_{\vz \in K} \va \vz}}$ to denote applying the CG cut
induced by $\va$ to $K$. We may extend this to an ordered list $\CL =
(\va_1,\dots,\va_k)$, letting $\CG(K,\CL)$ be the result of applying the CG cuts
induced by $\va_1,\dots,\va_k$ to $K$ one by one in this order (from left to
right).   

In terms of certifying such cuts, if $K = \set{\vx \in \R^n: \MC \vx \leq \vd}$,
$\MC \in \Q^{m \times n}, \vd \in \Q^m$, is a polyhedron, then by Farkas's
lemma, every CG cut can be obtained as a conic combination of the constraints
after rounding down the right hand side. That is, for each $\vLa \geq 0$ such
that $\vLa \MC \in \Z^n$, we have the corresponding CG cut $\vLa \MC \vx \leq
\floor{\vLa \vd}$, and every CG cut for $K$ can be derived in this way. 

A cutting plane proof (CP) of integer infeasibility for $K \subseteq \R^n$ can
now be described as a list $\CL = (\va_1,\dots,\va_N)$, $\va_i \in \Z^n$, such
that $\CG(K,\CL) = \emptyset$. In this context, the number of CG cuts $N$
denotes the length of the CP proof. When $K = \set{\vx \in \R^n: \MC \vx \leq
\vd}$ is a polyhedron as above, to get a certified proof, we can augment $\CL$
with multipliers $\vLa_1 \in \R^m_+,\vLa_2 \in \R^{m+1}_+,\dots,\vLa_{N+1} \in
\R^{m+N}_+$ (we still refer to the length of $\CL$ as $N$ in this case).
Letting $\CL_i := (\va_1,\dots,\va_i)$, $i \in [N]$, the multipliers $\vLa_i \in
\R^{m+i-1}_+$, $0 \leq i \leq N$, certify the cut $\va_i \vx \leq \floor{\sup
\set{\va_i \vz: \vz \in \CG(K,\CL_{i-1})}}$, in the manner described in the
previous paragraph, using the original inequalities $\MC \vx \leq \vd$ (the
first $m$ components of $\vLa_i$) and the previous cuts $\va_j \vx \leq
\floor{\sup \set{ \va_j \vz: \vz \in \CG(K,\CL_{j-1})}}$, $j \in [i-1]$.
Finally, $\vLa_{N+1} \in \R^{m+N}_+$ provides the Farkas certificate of
infeasibility for $\CG(K,\CL)$, using the original system together with all the
cuts. 

As with branching proofs, it is important to be able to control the bit-size
$\inner{\CL}$ of a CP proof and not just its length (i.e., the number of cuts in
the list $\CL$). Here $\inner{\CL}$ corresponds to the number of bits needed to
describe $\inner{\va_1,\dots,\va_N}$, as well as $\inner{\vLa_1,\dots,\vLa_{N+1}}$
for a certified proof. When $K = \set{\vx \in \R^n: \MC \vx \leq \vd}$ is a
polyhedron as above, a fundamental theorem of Cook, Coullard and
Tur\'{a}n~\cite{CCT87} is that any CP proof $\CL$ of integer infeasibility for
$K$ can be recompiled into a \emph{certified} CP proof $\CL'$, such that $N :=
|\CL| = |\CL'|$ and $\inner{\CL'} = \poly(N,L)$, where $L := \inner{\MC,\vd}$ is
the number of bits needed to describe the linear system defining $K$. Thus, for
CP proofs on polyhedra, one can, without loss of generality, assume that the
bit-size of a CP proof is polynomially related to its length and the bit-size of
the defining linear system. 

In terms of general complexity upper bounds, another important theorem
of~\cite{CCT87} is that every integer free rational polytope $K \subseteq \R^n$
admits a CP proof of infeasibility of length $O(f(n)^n)$, where $f(n)$ is the
flatness constant. This bound was achieved by showing that a run of Lenstra's
algorithm can effectively be converted into a CP proof.  

To relate CP and branching proofs, there is a simple disjunctive characterization of
CG cuts. Namely, $\va \vx \leq b$, $\va \in \Z^n$, $b \in \Z$ is a CG cut for $K$
iff $\set{\vx \in K: \va \vx \geq b+1} = \emptyset$. That is, if and only if the
\emph{right side} of the disjunction $\va \vx \leq b$ or $\geq b+1$ is empty for
$K$. From this observation, one can easily show that any CP proof of infeasibility
can be converted into a branching proof of infeasibility with only an $O(1)$
factor blowup in length (see~\cite{BFIKPPR18} for a formal proof). 

\subsection{Complexity of Branching Proofs} 

Despite its long history of study within IP, general branching has only recently
been studied from the SAT perspective. In~\cite{BFIKPPR18}, Beame et al rediscovered the concept of general branching proofs in the context of SAT, naming
them \emph{stabbing planes} (SP) refutations, and analyzed them from the proof
complexity perspective. To keep with this nomenclature, we use the term stabbing
planes (SP) refutations to refer specifically to a certified branching proofs of
infeasibility for SAT formulas. In terms of results, they showed that SP
refutations can size- or depth-simulate CP proofs and showed that they are equivalent to Kraj{\'\i}{\v{c}}ek's~\cite{K98a} tree-like R(CP) refutations.
They further gave lower bounds and impossibility results, showing an
$\Omega(n/\log n)$ lower bound on the depth of SP refutations and showed that SP
refutations cannot be balanced. 

Lastly, they provided upper bounds on the length of SP refutations, showing that
any Tseitin formula has a quasi-polynomial sized SP refutation. We recall that a
Tseitin formula is indexed by a constant degree graph $G = (V,E)$ and a set of parities $l_v \in \{0,1\}$, $v \in V$ satisfying $\sum_{v \in V} l_v \equiv 1
\mod 2$. The variables $\vx \in \set{0,1}^E$ index the corresponding subset of
edges, where the assignment $\vx$ is a satisfying assignment iff $\sum_{e \in E:
v \in e} x_e \equiv l_v \mod 2$, $\forall v \in V$. Note that such a formula is
clearly unsatisfiable, since the sum of degrees of any (sub)graph is even
whereas $\sum_{v \in V} l_v$ is odd by assumption. For such formulas, Beame et
al gave a $2^\Delta (n \Delta)^{O(\log n)}$ length SP refutation, where
$\Delta$ denotes the maximum degree of $G$. A long standing
conjecture~\cite{Beame00,BFIKPPR18} is that Tseitin formulas are hard for
cutting planes, and the above result was seen as evidence that SP refutations are
strictly stronger than CP. We note that exponential lower bounds for CP were
first proven by Pudl{\'a}k~\cite{Pudlak97}, who showed how to derive CP lower
bounds from monotone circuit lower bounds. However, the corresponding monotone circuit
problem for Tseitin formulas is easy, and hence cannot be used for proving strong lower bounds.  

Beame et al~\cite{BFIKPPR18} left open some very natural proof complexity theoretic
questions about branching proofs, which highlighted fundamental gaps in our
understanding of the proof system. Their first question relates to the
relationship between bit-size $\inner{\CT}$ and length $|\CT|$ of an SP proof.
Precisely, they asked whether one can always assume that the bit size of an SP
refutation is bounded by a polynomial in the dimension and the length of the proof. That is, can an SP refutation be ``recompiled'' so that it satisfies this
requirement without increasing its length by much? As mentioned previously, the
corresponding result for CP refutations was already shown by Cook et
al~\cite{CCT87}, though the techniques there do not seem to apply to SP. Their
second question was whether one could show a separation between CP and SP, which
would follow if Tseitin formulas are (say, exponentially) hard for CP. Lastly,
they asked whether one can prove super-polynomial lower bounds for SP.

\subsection{Our Contributions}

In this work, we give answers to many of the questions above. Firstly, we
resolve Beame et al's bit-size vs length question affirmatively. Secondly, we
show that Tseitin formulas have quasi-polynomial size CP proofs, showing that
they do not provide an exponential separation between CP and SP. Lastly, we give a very
simple family of $n$-dimensional (mixed-)integer free polytopes for which any
branching proof has size exponential in $n$. We describe these contributions in
detail below.

\subparagraph*{Bit-size of Branching Proofs} As our first main contribution, we
resolve Beame et al's bit-size vs length question, by proving the following more
general result:

\begin{theorem}
\label{thm:coef-bnd} 
Let $K \subseteq \R^n$ be an integer free compact convex set satisfying $K
\subseteq R\B_1^n$, where $\B_1^n$ is the $\ell_1$ ball and $R \in \N$. Let $\CT$
be a branching proof of integer infeasibility for $K$. Then, there exists a
branching proof $\CT'$ for $K$, such that $|\CT'| \leq O(n |\CT|)$, and where
every edge $e$ of $\CT'$ is labeled by an inequality $\va_e' \vx \leq b_e'$,
$\va_e' \in \Z^n$, $b_e' \in \Z$, and $\max \set{\|\va_e'\|_\infty, |b_e'|}
\leq (10nR)^{(n+2)^2}$. Moreover, $\inner{\CT'} = O(n^3 \log_2(2nR) |\CT|)$.  
\end{theorem}

The above theorem says that, at the cost of increasing the number of nodes in the branching tree by a factor $O(n)$, one can reduce the coefficients in the
normals of the disjunctions to $(10nR)^{(n+2)^2}$. In particular, since
$\va'_e,b'_e$ are integral, they can be described with $O(n^3\log_2(2nR))$ bits.
We note that the final bound on $\inner{\CT'}$ ends up being better than $O(n^3
\log_2(2nR)|\CT'|) = O(n^4 \log_2(2nR)|\CT|)$, due to the fact that the ``extra''
nodes we need in $\CT'$ use smaller disjunctions that are describable using
$O(n^2 \log_2(2nR))$ bits. In the context of SAT, the desired bound on the coefficients
of SP proofs follows directly from the fact that any SAT polytope, as
in~\eqref{SAT-LP}, is contained inside $[0,1]^n \subseteq n \B_1^n$.

As mentioned previously, one would generally want a branching
proof to come with certificates of infeasibility for the leaf nodes. For a
rational polytope $K$, the following corollary bounds the cost of extending the
branching proof produced by Theorem~\ref{thm:coef-bnd} to a certified branching proof. To be precise, the bit-size of the final certified proof can be made proportional to the size of the original tree, the bit-encoding length of the
defining system for $K$ and a polynomial in the dimension. 

\begin{corollary} 
\label{cor:coef-bnd-pol}
Let $K = \{\vx \in \R^n: \MC \vx \leq \vd\}$ be rational polytope with $\MC \in
\Q^{m \times n}, \vd \in \Q^m$ having bit-size $L := \langle \MC,\vd \rangle$. Let
$\CT$ be a branching proof for $K$. Then there exists a certified branching
proof $\CT'$ for $K$ such that $|\CT'| \leq O(n) |\CT|$ and $\langle \CT'
\rangle = O(n^6 L) |\CT|$.
\end{corollary}

The bit-size $L := \inner{\MC,\vd}$ of $K$ in Corollary~\ref{cor:coef-bnd-pol}
shows up for two related reasons. Firstly, we need $L$ to upper bound the
$\ell_1$ circumradius $R$ of $K$, which is in turn used to bound the bit-size of
the disjunctions in Theorem~\ref{thm:coef-bnd}.  For a rational polytope $K$,
$R$ is in fact always upper bounded by $2^{O(L)}$. We stress that $2^{O(L)}$
more directly upper bounds the $\ell_1$ norm of the vertices of $K$, which in
turns upper bounds the $\ell_1$ circumradius of $K$ only under the assumption
that $K$ is indeed bounded (i.e., that $K$ is polytope and not just a
polyhedron). However, it is well known that for a rational polyhedron $K$, $K
\cap \Z^n = \emptyset$ iff $K \cap 2^{O(L)} \B_1^n \cap \Z^n = \emptyset$ (see
Schrijver~\cite{Schrijver86} Chapter 17). Therefore, the boundedness assumption
above is essentially without loss of generality. More precisely, one can simply
add box constraints $-2^{O(L)} \leq x_i \leq 2^{O(L)}$, $i \in [n]$, to the description of $K$, which increases the description length by $O(n)$. 
The second reason for needing $L$ is to bound the bit-complexity of the Farkas
infeasibility certificates at the leaves of the modified branching tree. By
standard bounds, such a certificate has bit-size bounded by $O(n)$ times the bit
description length of a minimal infeasible subsystem (over the reals) at the
corresponding leaf. By Helly's theorem, a minimal infeasible subsystem has at
most $n+1$ inequalities consisting of a subset of the inequalities defining $K$
and the inequalities from branching, where each of these inequalities has
bit-size at most $O(n^3 L)$ by Theorem~\ref{thm:coef-bnd}.  

\subparagraph*{Sketch of Theorem~\ref{thm:coef-bnd}} We now give some intuition
about the difficulties in proving Theorem~\ref{thm:coef-bnd}, which is
technically challenging, and sketch the high level proof ideas. 

We first note that any disjunction $\va \vx \leq b$ or $\geq b+1$, where $\va$ has
very large coefficients, only cuts off a very thin slice of $K$. In particular,
the width of the band $b \leq \va \vx \leq b+1$ is exactly $1/\|\va\|_2$. Thus, it
is perhaps intuitive that any ``optimal'' proof should use wide disjunctions
instead of thin ones, and hence should have reasonably small coefficients.
Unfortunately, this intuition turns out to be false. Indeed, disjunction angles 
can be more important than their widths for obtaining proofs of optimal length. 

The following simple 2 dimensional example shows that if one wishes to exactly
preserve the length of a branching proof, then large coefficients are unavoidable
even for sets of constant radius. Examine the line segment
\[
K = \set{(x_1,x_2): M x_1 + x_2 = 1/2, 0 \leq x_2 \leq 2},
\]
for $M \geq 1$. Clearly, branching on $M x_1 + x_2 \leq 0$ or $\geq 1$ certifies
integer infeasibility in one step. Now let $\va \in \Z^2$ be any branching
direction that also certifies infeasibility in one step. Then, the width of $K$
with respect to $\va$ must be less than one:
\[
\max_{\vx \in K} \va \vx - \min_{\vx \in K} \va \vx = |2(a_2 - a_1/M)| < 1.
\]
Now if $a_2 \neq 0$, then $|a_1| \geq M/2$, so $\|\va\|_\infty \geq M/2$. If
$a_2 = 0$, then we should let $\va = (1,0)$, since this choice yields the widest
possible disjunctions under this restriction. Branching on $\va = (1,0)$ cannot
certify infeasibility in one step however, since $\vx = (0,1/2) \in K$ and $\va
\vx = 0$.  

To recompile a proof $\CT$ using only small coefficients, we must thus 
make do with a discrete set of disjunction angles that may force us to
increase the length of the proof. Given an arbitrary
branching direction $\va$, the standard tool for approximating the direction of
$\va$ using small coefficients is so-called \emph{Diophantine} approximation
(see Lemma~\ref{lem:diophantine}). Thus, the natural first attempt would be to
take every disjunction $\va \vx \leq b$ or $\geq b+1$ in $\CT$ and replace it by
its small coefficient Diophantine approximation $\va' \vx \leq b'$ or $\geq b'+1$
to get $\CT'$. As shown above, there are examples where any such small coefficient $\CT'$ will no longer be valid, due to some of the leaf nodes
becoming feasible. 

Let $v \in \CT$ be a leaf node with relaxation $K_v = \set{\vx \in K: \MA\vx \leq
\vb} = \emptyset$ and corresponding approximation $v' \in \CT'$ with $K_{v'} =
\set{\vx \in K: \MA'_v\vx \leq \vb'_v} \neq \emptyset$. To transform $\CT'$ to a
valid proof, we must therefore add branching decisions to $\CT'$ below $v'$ to
certify integer-freeness of $K_{v'}$. From here, the main intuitive observation
is that since $P_v := \MA_v \vx \leq \vb_v$ and $P_{v'} := \MA'_v \vx \leq \vb'_v$ have
almost the same inequalities, $P_{v'} \cap K$ should be very close to infeasible.

By inspecting a Farkas-type certificate of infeasibility of $K \cap P_v$ (see
subsection~\ref{subsec:gen-farkas}), for a good enough Diophantine approximation
$P_{v'}$ to $P_v$, one can in fact pinpoint an inequality of $P_{v'}$, say
$\va'_{v,1}
\vx \leq b'_{v,1}$, such that replacing $b'_{v,1}$ by $b'_{v,1}-1$ makes $K \cap P_{v'}$
empty. This uses the boundedness of $K$, i.e., $K \subseteq R\B_1^n$, and that
the disjunctions induced by the rows of $\MA'$ are much wider than those induced
by $\MA$. Note that the emptiness of $\va'_{v,1} \vx \leq b_{1,v}'-1$ corresponds to saying
that $\va'_{v,1} \vx \geq b'_{v,1}$ is a valid CG cut for $K \cap P_{v'}$. Furthermore,
this CG cut has the effect of reducing dimension by one since now $\va'_{v,1}
\vx = b'_{v,1}$.

Given the above, it is natural to hope than one can simply repeat the above
strategy recursively. Namely, at each step, we try to find a new CG cut induced
by a row of $\MA'$ which reduces dimension of $K \cap P_{v'}$ by one. Unfortunately, the strategy as stated breaks down after one step. The main problem is that, after the first step, we have no ``information'' about
$\va_{v,1}
\vx \leq b_{v,1}$ restricted to $\va'_{v,1} \vx = b'_{v,1}$. Slightly more precisely, we no
longer have a proxy for $\va_{v,1} \vx \leq b_{v,1}$ in $P_{v'}$ that allows us to push
this constraint ``backwards'' on the subspace $\va'_{v,1} \vx = b'_{v,1}$. Since we must
somehow compare $P_{v'}$ to $P_v$ to deduce infeasibility, this flexibility turns
out to be crucial for being able to show the existence of a dimension reducing
CG cut.

To fix this problem, we rely on a more sophisticated iterated form of
Diophantine approximation due to Frank and Tardos~\cite{FT87}. At a high level
(with some simplification), for a disjunction $\va \vx \leq b$ or $\geq b+1$, $\va \in \Z^n$, $b \in \Z$, we first construct a sequence of Diophantine
approximations $\va_1,\dots,\va_k \in \Z^n$, containing $\va$ in their span,
which intuitively represents the highest to lower order bits of the direction of
$\va$. From here, we carefully choose a sequence $b_1,\dots,b_k \in \Z$ indexing
inequalities $\va_i \vx \leq b_i$, $i \in [k]$, which allows us to get better and
better approximations of $\va \vx \leq b$. Since we are, in reality, replacing the disjunction $\va \vx \leq b$ or $\geq b+1$, we will in fact need a sequence that
somehow approximates both sides of the disjunction at the same time. This will
correspond to requiring that a ``flipped'' version of the sequence, namely
$\va_i \vx \geq b_i$, $i \in [k-1]$, and $\va_k \vx \geq b_{k}+1$, gives
improving approximations of $\va \vx \geq b+1$. Restricting attention to just
the $\va \vx \leq b$ side, we will show the existence of improving ``error
levels'' $\gamma_1 \geq \gamma_2 \geq \dots \geq \gamma_k = 0$, such that
$\|\vx\|_1 \leq R, \va_l \vx \leq b_l, \va_i \vx = b_i, i \in [l-1] \Rightarrow
\va \vx \leq b+\gamma_l$. Furthermore, we will ensure that branching on
$\va_l \vx \leq b_l-1$, not only reduces the error bound
$\alpha_l$, but in fact implies a far stronger inequality than $\va \vx \leq b$.
Precisely, we will require $\|\vx\|_1 \leq R, \va_l \vx \leq b_l-1, \va_i \vx = b_i, i \in [l-1] \Rightarrow
\va \vx \leq b-n\gamma_l$. Hence, once we have learned the equalities $\va_i \vx
= b_i$, $i \in [l-1]$, $\va_l$ becomes a suitable proxy for
$\va$ which we can use to push the constraint $\va \vx \leq b$ ``backwards''. 
Note that if $l=k$, we have in fact fully learned $\va \vx
\leq b$ since $\alpha_k = 0$. If $l < k$ and $\va \vx \leq b$ is
the ``closest inequality to infeasibility'' in the current relaxation,
corresponding to the inequalities in $P_{v'}$ for some leaf $v'$ together with
the additional equalities as above, we will be able to guarantee that the CG
cuts induced by $\va_l$ and $-\va_l$ yield the new equality $\va_l \vx = b_l$.
Note that if we always manage to reduce dimension by at least $1$, we will
terminate with an infeasible node after adding at most $n+1$ pairs of CG cuts.
So far, we have discussed replacing a disjunction $\va \vx \leq b$ or $\geq b+1$ by a sequence instead of a single disjunction, and the latter is what is actually
needed. For this purpose, the new disjunction will have the form $\va' \vx \leq
b'$ or $\geq b'+1$ where $\va' = \sum_{i=1}^k M^{k-i} \va_i$ and $b' =
\sum_{i=1}^k M^{k-i} b_i$ for $M$ chosen large enough. This is chosen to ensure
that $\|\vx\|_1 \leq R, \va' \vx \leq b', \va_i \vx = b_i, i \in [l-1]$ ``almost
implies'' $\va_l \vx \leq b_l$, with a symmetric guarantee for the flipped
sequence. The full list $(a',b',k,a_1,b_1,\gamma_1,\dots,a_k,b_k,\gamma_k)$
is what we call a \emph{valid substitution
sequence} of $\va \vx \leq b$ (see definition~\eqref{def:validDioph}). The main
difficulty in constructing and analyzing the disjunction $\va' \vx \leq b'$ or
$\geq b'+1$, is that each side of the disjunction should induce a valid
substitution sequence for the corresponding side of $\va \vx \leq b$ or $\geq
b+1$. That is, we need to work for ``both sides'' at once. As the remaining details are technical, we defer further discussion of the proof to
Section~\ref{sec:coefficients} of the paper. 

As a point of comparison, we note that in contrast to
Theorem~\ref{thm:coef-bnd} the recompilation result of~\cite{CCT87} does not
give a \emph{length independent bound} on the size of normals of the CG cuts it
produces (e.g., depending only on the $\ell_1$ radius of $K$). An interesting
question is whether one can give length independent bounds for CP proofs based
only on the bit-complexity $L$ of the starting system. Perhaps one avenue for
such a reduction would be to first convert the CP proof to a branching proof
and try to apply the techniques above. The main issue here is that the first reduction phase above, which approximates each disjunction in the tree with a
small coefficient one, need not preserve the CP structure. Namely, after the
replacement, it is not clear how to guarantee that every disjunction in the
replacement tree has at least one ``empty'' side (note that this problem is
compounded by the approximation errors going up the tree).

\subparagraph*{Upper Bounds for Tseitin formulas} As our second contribution,
we show that Tseitin formulas have quasi-polynomial CP proofs, refuting the
conjecture that these formulas are (exponentially) hard for CP. 

\begin{theorem} 
\label{thm:tseitin}
Let $G = (V,E)$ be an $n$-vertex graph, $l_v \in \{0,1\}$, for $v \in V$, be parities
and $\Phi$ be the corresponding Tseitin formula. Then $\Phi$ has a CP refutation
of length $2^\Delta (n \Delta)^{O(\log n)}$, where $\Delta$ is the maximum
degree of $G$.
\end{theorem}

To prove the theorem our main observation is that the quasi-polynomial SP
proof of Beame et al~\cite{BFIKPPR18} is of a special type, which we dub an
\emph{enumerative branching proof}, that can be automatically converted to a CP
proof of at most twice the length.

We define an \emph{enumerative} branching proof for a compact convex set $K$ to
correspond, as before, to a tree $\CT$ with root $r$ and root relaxation $K_r :=
K$. At every node $v \in \CT$ with $K_v \neq \emptyset$, we choose a branching
direction $\va_v \in \Z^n \setminus \set{0}$ and immediately branch on all
possible choices $b \in \Z$ that intersect the current relaxation $K_v$. Note
that tree $\CT$ need no longer be binary. Formally, we first label $v$ with the
bounds $l_v, u_v \in \R$ satisfying 
\[
\set{\va_v \vx: \vx \in K_v} \subseteq [l_v,u_v].
\]
From here, we create a child node $v_b$, for every $b \in \Z$ such that $l_v
\leq b \leq u_v$. The edge $e = \{v,v_b\}$ is now labeled with the
\emph{equality} $\va_v \vx = b$ and the updated relaxation becomes $K_{v_b} =
\set{\vx \in K_v: \va_v \vx = b}$. From here, each leaf node $v \in \CT$ can
be of two different types. Either $K_v = \emptyset$, or if $K_v \neq \emptyset$,
the interval $[l_v,u_v]$ is defined and does not contain integer points, i.e.,
$\floor{u_v} < l_v$. A tree $\CT$ satisfying the above
properties is a valid enumerative branching proof of integer infeasibility for
$K$. 

It is an easy exercise to check that any enumerative branching proof can be
converted to a standard branching proof incurring only a constant factor blowup
in the number of nodes. Theorem~\ref{thm:tseitin} follows directly from the
observation that the Beame et al SP proof is enumerative together with the
following simulation result.   

\begin{theorem}
\label{thm:branching-to-cp}
Let $K \subseteq \R^n$ be a compact convex set. Let $\CT$ be an
enumerative branching proof of $K$. Then there exists $\CL = (\va_1,\dots,\va_N)
\in \Z^n$ such that $\CG(K,\CL) = \emptyset$ and $N \leq 2|\CT|-1$.
\end{theorem}

While in the above generality the result is new, the main ideas (at least for
rational polytopes) are implicit in Cook et al~\cite{CCT87}. In particular,
their proof that any integer free rational polytope admits a CP proof of length
at most $O(f(n)^n)$ in effect treats Lenstra's algorithm as an enumerative
branching proof which they serialize to get a CP proof.
Theorem~\ref{thm:branching-to-cp} shows that their serialization technique is
fully general and in fact can be applied to any enumerative branching proof.  To
get a certified CP proof of small bit-size from
Theorem~\ref{thm:branching-to-cp} for a rational polyhedron $K$, we note that it
suffices to apply the recompilation technique of Cook et al~\cite{CCT87} to the
output of Theorem~\ref{thm:branching-to-cp}.  While there is some technical
novelty in the generalization to arbitrary compact convex sets, we feel the main
contribution of Theorem~\ref{thm:branching-to-cp} is conceptual. As evidenced
by Theorem~\ref{thm:tseitin}, the formalization of enumerative branching proofs
and their relationship to CP can be a useful tool for constructing CP proofs. 


We now sketch the main ideas for serializing an enumerative branching proof
$\CT$ for $K$. We start from the root $r \in \CT$, with branching direction
$\va_r \in \Z^n$ and $\set{\va_r \vx: \vx \in K} \subseteq [l_r,u_r]$. The idea
is to iteratively ``push'' the hyperplane $H_{b} = \set{\vx \in \R^n:
\va_r \vx = b}$, with $b$ initialized to $u_r$, backwards through $K$, until $K$ is empty (i.e.,
iteratively decreasing $b$ until it goes below $l_r$). The first push is given
by the CG cut induced by $\va_r$ which pushes $H_{u_r}$ to $H_{\floor{u_r}}$.
That is, $b \leftarrow \floor{b}$. Since $b$ is now integral, we can no longer
decrease $b$ just using CG cuts induced by $\va_r$.  At this point, we note that
the subtree $\CT_{r_b}$ of $\CT$ rooted at the child $r_b$ is a valid branching
proof for $K \cap H_b$. We can thus apply the procedure recursively on $K \cap
H_b$ and $\CT_{r_b}$ to ``chop off'' $K \cap H_b$. For this purpose, one
crucially needs to be able to lift CG cuts applied to the face $K \cap H_b$ to
CG cuts one can apply to $K$ that have the same effect on $K \cap H_b$. Such a
\emph{lifting lemma} is classical for rational polyhedra~\cite{Chvatal73} and
was established more recently for compact convex sets in~\cite{DDV14}, a variant
of which we use here. Applying the lifted CG cuts to $K$, we can thus guarantee
that $K \cap H_b = \emptyset$.  This allows us to push once more with the cut
induced by $\va_r$, pushing $H_b$ to $H_{b-1}$. The process now continues in a
similar fashion until $K$ is empty. We note that the enumerative structure is
crucial here, as it allows one to keep the ``action'' on the boundary of $K$
throughout the entire proof.  

\subparagraph*{Lower Bounds for Branching Proofs} As our final contribution, we
give a simple family of $n$-dimensional (mixed-)integer free polytopes which require
branching proofs of length exponential in $n$.

\begin{theorem} 
\label{thm:lower-bnd}
The integer-free SAT polytope 
$$P_n := \set{\vx \in [0,1]: \sum_{i \in S} x_i +
\sum_{i \not \in S} (1-x_i) \geq 1, \forall S \subseteq [n]}$$
requires branching
proofs of length $2^n/n$.  
\end{theorem}

The above example is due to Cook et al~\cite{CCT87}, which they used to give a
$2^n/n$ lower bound for CP. In the above theorem, we show that their lower bound
technique extends to branching proofs. As it is very simple and short, we give
the full proof below.

\begin{proof}
The first observation is that $P_n$ is ``integer critical'', namely,
removing any constraint from $P_n$ makes the polytope integer feasible. In
particular, removing $\sum_{i \in S} x_i + \sum_{i \not \in S} (1-x_i) \geq 1$,
for any $S \subseteq [n]$, makes the vector $\boldsymbol{1}_{\bar{S}}$, the
indicator of the complement of $S$, feasible.

Let $\CT$ denote any branching proof for $P_n$. For any leaf node $v$ of $\CT$,
by Farkas's lemma, the infeasibility of the continuous relaxation $(P_n)_v$ is
certified by at most $n+1$ constraints. Since $P_n$ is non-empty, at most $n$ of
these constraints can come from the description of $P_n$. Letting $N$ denote the
number of leaves of $\CT$, one can therefore certify the infeasibility of each
leaf of $\CT$ using at most $nN$ original constraints from $P_n$. If $nN <
2^n$, then $\CT$ would certify the integer infeasibility of $P_n$ with at least
one constraint removed. By integer criticality of $P_n$, this is impossible.
Therefore $|\CT| \geq N \geq 2^n/n$, as needed.     
\end{proof}

One notable criticism of the above example is that it already has $2^n$
constraints. Thus, the length of the proof is simply proportional to the initial
representation. Interestingly, $P_n$ has a very simple extended formulation in
$\R^{2n}$ requiring only $O(n)$ constraints. A direct computation
reveals that
\begin{align*}
P_n &= \set{\vx \in [0,1]^n: \|(x_1-1/2,\dots,x_n-1/2)\|_1 \leq n/2-1} \\
    &= \set{\vx \in [0,1]^n: \exists \vy \in [0,1]^n, \sum_{i=1}^n y_i \leq n/2-1, \pm(x_i-1/2) \leq y_i, i \in [n]}. 
\end{align*}

Combining the above with Theorem~\ref{thm:lower-bnd}, we immediately get an
exponential lower bound for proving the mixed-integer infeasibility of a
compactly represented polytope. We note that in this setting, the lower bound is
indeed exponential in the description length of $P$.  

\begin{corollary} 
Let $Q_n = \set{(\vx,\vy) \in [0,1]^{2n}: \sum_{i=1}^n y_i \leq n/2-1,
\pm(x_i-1/2) \leq y_i, i \in [n]}$. Then any branching proof of mixed-integer
infeasibility for $Q_n$, proving $Q_n \cap \Z^n \times \R^n = \emptyset$,
has length at least $2^n/n$. 
\end{corollary}

To see the above, recall that a mixed-integer branching proof for $Q_n$ only
branches on integer disjunctions supported on the first $n$ variables. Thus, it
is entirely equivalent to a branching proof for the projection of $Q_n$ onto these
variables, namely, to a branching proof for $P_n$. 

As a final remark, we note that in the extended space, $Q_n$ does in fact have a
very short proof of infeasibility using only $n$ split cuts, which are perhaps
the most important class of cutting planes in practice (in fact, the most
generically effective cuts are the Gomory mixed-integer cuts (GMI), which are
equivalent to split cuts for rational polyhedra~\cite{CL01}). Roughly speaking,
a split cut here is any linear inequality that is \emph{valid for both sides}
$\va \vx \leq b$ or $\geq b+1$, $\va \in \Z^n$, $b \in \Z$, of an integer
disjunction. In particular, $y_i \geq 1/2$ is a valid split cut for $Q_n$, for
$i \in [n]$, since it is valid for $x_i \leq 0$ and $x_i \geq 1$.  These $n$
splits together imply that $\sum_{i=1}^n y_i \geq n/2$, and thus adding them to
$Q_n$ makes the system infeasible. 
 
\subsection{Conclusions}

In this work, we have continued the proof complexity theoretic study of
branching proofs started in~\cite{BFIKPPR18}, establishing analogues of the CP
results in~\cite{CCT87} for branching proofs. In the process, we have clarified
basic properties of the branching proof system, including how to control the
size of coefficients, how to simulate important classes of branching proofs
using CP, and how to construct elementary lower bound examples for them. We
hope that these results will help motivate a further study of this important
proof system.  

In terms of open questions, there are many. A first question is whether size of
the coefficients in Theorem~\ref{thm:coef-bnd} can be reduced from
$(nR)^{O(n^2)}$ to $(nR)^{O(n)}$. The latter corresponds to an upper bound
on the coefficients of an integer hyperplane passing through $n$ integer points
in $[-R,R]^n$, and is also a natural from the perspective of Diophantine
approximation. We note that the $(nR)^{O(n^2)}$ dependency is due to the form
$\va' = \sum_{i=1}^k M^{k-i} \va_i$ of the approximating disjunctions, where we
need $M = (nR)^{O(n)}$ to ensure that the different levels present in $\va'$
don't ``interfere'' with each other. On the lower bound side, in the context of
SAT, the example we use has exponentially many clauses. It would be much more
interesting to find polynomial sized formulas with exponential sized branching
proofs. In the context of integer programming, as mentioned previously, the best
known algorithms for general integer programming require $n^{O(n)}$ time. A very
interesting question is whether one can find an example of an integer free
compact convex set $K \subseteq \R^n$, requiring branching proofs of size
$n^{\Omega(n)}$. Such a lower bound would show that Lenstra-type algorithms for
IP, which in fact yield enumerative branching proofs, cannot be substantially
improved. We note that this still leaves open the possibility that so-called
Kannan-type algorithms can do much better (see~\cite{thesis/D12} Chapter 7 for a
reference). In terms of upper bounds, a natural question is whether one can
leverage the simulation of enumerative branching proofs by CP to give new upper
bounds beyond Tseitin formulas. It was shown by Cook et al~\cite{CCT87} that for
SAT, CP can be simulated by extended resolution. A natural question is whether
stabbing planes can also be simulated by extended resolution. Lastly, as
mentioned previously, it would be interesting to establish length independent
bounds for the coefficients of the normals in CP proofs.

\subsection{Acknowledgments}

The first author would like to deeply thank Noah Fleming, Denis Pankratov, Toni
Pitassi and Robert Robere for posing the bit-size vs length question for SP and
for very stimulating conversations while the author was visiting the University
of Toronto. The authors are also very grateful for the comments from the anonymous
reviewers, which have greatly helped us improve the quality of the presentation.

\subsection{Organization}

In Section~\ref{sec:prelims}, we collect basic notation, formalize the
definition of branching proofs and cover the necessary tools from Diophantine
approximation. In Section~\ref{sec:coefficients}, we present our branching proof
recompilation theorem, which ensures that the bit-size of branching proofs can
be polynomially bounded. In Section~\ref{sec:simulation}, we show how to
simulate enumerative branching proofs via CP, and apply this simulation to get
a quasi-polynomial CP bound for Tseitin formulas. 

\section{Preliminaries}
\label{sec:prelims}

\subparagraph*{Basic Notation} The natural numbers are denoted by $\N$, the
reals and non-negative reals by $\R,\R_+$ respectively. For $m \in \N$, we
denote the set $\set{1,\dots,m}$ by $[m]$. Vectors $\vx \in \R^n$
are denoted in bold and scalars by $x \in \R$. The standard basis vectors of $\R^n$ are denoted by $\ve_i, i \in [n]$. Given two vectors $\vx,\vy \in
\R^n$, we write $\vx \vy := \sum_{i=1}^n x_i y_i$ for their inner product. The
$\ell_1$ and $\ell_\infty$ norm of $\vx$ are $\|\vx\|_1 = \sum_{i=1}^n |x_i|$
and $\|\vx\|_\infty = \max_{i \in [n]} |x_i|$ respectively.  We denote the
$\ell_1$ ball in $\R^n$ by $\B_1^n = \set{\vx \in \R^n: \|\vx\|_1 \leq 1}$. For a
vector $\vx = (x_1,\dots,x_n) \in \R^n$, we let $\round{\vx} :=
(\round{x_1},\dots,\round{x_n})$ denote the vector whose coordinates are those
of $\vx$ rounded to the nearest integer. 

Since we shall study convex bodies lying in the $l_1$ ball of some radius $R \in \N$, it is helpful to define the following shorthand notation: for a set of linear inequalities $\MA \vx  \leq \vb $ and a vector $\vc$, the expression $\MA \vx  \leq \vb  \rimplies \vc \vx  \leq d$ stands for
$$\set{\vx \in \R^n : \length{\vx }_1 \leq R, \MA \vx  \leq \vb } \subseteq
\set{ \vx \in \R^n : \length{\vx }_1 \leq R, \vc \vx  \leq d }.$$

\begin{definition}[Halfspace, Hyperplane]
For $\va \in \R^n$, $b \in \R$, we define the halfspace $H_{\va,b} = \set{\vx \in
\R^n: \va \vx \leq b}$ and the hyperplane $H^=_{\va,b} = \set{\vx \in \R^n: \va
\vx = b}$.
\end{definition}

\begin{definition}[Support Function]
Let $K \subseteq \R^n$. The \emph{support function} $h_K:
\R^n \rightarrow \R$ is defined as $h_K(\va) := \sup_{\vx \in K} \va \vx$. The support
function is always convex and is continuous if $K$ is non-empty and bounded. If
$K$ is non-empty and compact, the supremum in $h_K(\va)$ is always attained. By
convention, if $K = \emptyset$ we define $h_K(\va) = -\infty$, $\forall \va \in \R^n$.  

For $K \subseteq \R^n$ non-empty and compact and $\va \in \R^n$, we define the supporting
hyperplane of $K$ induced by $\va$ to be $H^=_K(\va) := \set{\vx \in \R^n: \va
\vx = h_K(\va)}$. We define the set of maximizers of $\va$ in $K$ to be
$F_K(\va) := K \cap H^=_K(\va)$. 
\end{definition}

\subsection{Bit-Sizes}

\begin{definition}[Bit-size]\label{def:bit-size} The notation $\inner{x}$ is reserved for the number of bits required to express the object $x$, or the bit-size of $x$. We build up the precise definitions as follows:

For $r \in \Q, r = p/q, p \in \Z, q\in \Z, q>0, \inner{r} := 1 + \ceil{\log_2
(|p| + 1)} + \ceil{\log_2 (q + 1)}$. Next, for $\vc \in \Q^n$ with $\vc = (c_1, c_2 \ldots c_n), \inner{c}:= n + \sum_{i=1}^n \inner{c_1}$. Similarly for matrices $\MA \in \Q^{m \times n}$, $\inner{\MA}:= mn + \sum_{i =1}^m \sum_{j=1}^n \inner{\MA_{ij}}$. $\inner{A, B}$ is simply $\inner{A} + \inner{B}$ when these terms are well-defined. 

For a labeled rooted tree $\CT$ with $n$ nodes and $m$ edges $E[\CT]$, and where edges $e \in E[\CT]$
have labels $L_e$ and nodes $v$ have labels $L_v$, and if the labels belong to a
class for which the bit-size has already been defined, then $\inner{\CT} := n +
m +
\sum_{e \in E[\CT]} \inner{L_e} + \sum_{v \in \CT} \inner{L_v}$. 

\end{definition}

\subsection{Branching Proofs}
\begin{definition}[Branching Proof]
A branching proof of integer infeasibility for a convex set $K \subseteq \R^n$
is represented by a rooted binary tree $\CT$ with root $r := r_\CT$. Each
node $v \in \CT$ is labeled with $(\va_v, b_v), \va_v \in \Z^n, b_v \in \Z$ and has two children nodes: the left child $v_l$ and right child $v_r$. Since the inner product of two integer vectors is an integer, the integer lattice $\Z^n$ can be partitioned into $\set{ \vx \in \Z^n: \va_v \vx \leq b_v}, \set{\vx \in \Z^n: \va_v \vx \geq b_v +1}$. This partition is referred to as the branch or integer disjunction given by $(\va_v, b_v)$.

Every edge $e \in E[\CT]$ is labeled with an inequality $\va_e \vx \leq b_e$. A left edge $e_l = \set{v, v_l}$ is labeled with $\va_v \vx \leq b_v$, or that $\va_e = \va_v, b_e = b_l$. However, a right edge $e_r = \set{v,v_r}$ is labeled with $\va_v \vx \geq b_v +1$, so that $\va_e = -\va_v, b_e = -b_1 -1$.

For each node $v \in \CT$, we define
$P_{\CT}(v)$ to be the unique path from the root $r$ of
$\CT$ to $v$. Also define for each $v$ a polyhedron $P_v = \set{\vx \in \R^n :
\MA_v \vx \leq \vb_v}$ where the rows of $\MA_v$ are given by $\va_{v, e}, e \in E[P_{\CT}(v)]$, and the coordinates of $\vb_v$ are $b_{v, e}, e \in E[P_{\CT}(v)]$. Let $K_v := K \cap P_v$. Note that $K_r = K$. 

For $\CT$ to be a proof of integer
infeasibility for $K$, we require that every leaf $v \in \CT$ ($v$ is a leaf if
its has no children) satisfies $K_v = \emptyset$.

We denote the length of the branching proof by $|\CT|$, which is defined to be
the number of nodes of $\CT$. The size of a branching proof $\inner{\CT}$ is simply its bit-size as a labeled rooted tree as given above in definition~\ref{def:bit-size}. 
\end{definition}

\begin{definition}[Certified Branching Proof] Suppose $K= \set{\vx \in \R^n :
\MC \vx \leq \vd}, \MC \in \Q^{r \times n}, \vd = \Q^r$ belongs to the
class of rational polyhedra. A certified branching proof of integer
infeasibility for $K $ is a standard branching proof $\CT$ of infeasibility of
$K$, but where every leaf node $v$ of $\CT$ is also labeled with a Farkas
certificate $\vLa_v \in \Q^{r+m_v}, \lambda_i \geq 0, \forall i \in [r+m_v]$,
where now $K_v = \set{\vx \in \R^n: \MC \vx \leq \vd, \MA_v \leq \vb_v}$, for
$\MA_v \in \R^{m_v \times n}, \vb_v \in \R^{m_v}$, $m_v = |P_\CT(v)|$. Let
$\vLa_v
= (\vLa_{v,1}, \vLa_{v,2}), \vLa_{v,1} \in \Q^r, \vLa_{v,2} \in \Q^{m_v}$. The requirement that
every $K_v = \emptyset$ for a leaf nodes $v$ is certified by requiring
$\vLa_{v,1}\MC + \vLa_{v,2}\MA_v = 0, \vLa_{v,1}\vd + \vLa_{v,2}\vb_v < 0$.

The bit-size of a certified branching proof is its bit-size when viewed as a labeled rooted tree.
\end{definition}

\begin{definition}[Enumerative Branching Proof] For a compact convex set $K$, an
\emph{enumerative branching proof} consists of a tree $\CT$ with root $r$ and
root relaxation $K_r := K$. Every node $v \in \CT$ is labeled with $(\va_v, l_v,
u_v)$, where $\va_v \in \Z^n, l_v, u_v \in \Q$ satisfying
$$\set{ \va_v \vx : \vx \in K_v} \subseteq [l_v, u_v].$$
There is a child of $v$ denoted $v_b$ for every $b \in \Z, l_v \leq b \leq u_v$, and the edge $e = \set{v, v_b}$ is labeled with the equality $\va_v \vx = b$. The relaxation at $K_{v_b}$ becomes $\set{ \vx \in K_v : \va_v \vx = b}$.

$\CT$ is a valid enumerative branching proof of infeasibility if every leaf node
$v \in \CT$ satisfies $K_v = \emptyset$ or $K_v \neq \emptyset$ but $[l_v, u_v]$
contains no integer points, i.e., $\floor{u_v} < l_v$.

$\inner{\CT}$ is again simply the bit-size of $\CT$ as a labeled rooted tree.
\end{definition}

\subsection{Simultaneous Diophantine Approximation}
 
The existence of a rational vector of small bit-size that well approximates an
arbitrary real vector is of prime importance in this paper. For this purpose, we
shall require standard tools from Diophantine approximation
(see~\cite{Schmidt80} for a reference). The following is a slightly adapted
version of Dirichlet's simultaneous approximation theorem, which will be
convenient for our purposes. We provide a proof for completeness.

\begin{lemma}
\label{lem:diophantine}
Let $\va  \in \R^n$ satisfy $\length{\va }_\infty = 1$ and let $N \geq 1$. Then, there exists a positive integer $l \leq N^n$ such that $\va' := \round{l\va}$ satisfies
$$\length{l\va  - \va '}_\infty < 1/N \quad \text{ and } \quad \|\va'\|_\infty = l \geq 1.$$ 
\end{lemma}
\begin{proof}
Let $\CC = \{I_\vz: \vz \in [N]^n\}$ denote the collection of $N^n$ half-open cubes forming a
partition of $[0,1)^n$, where $I_{\vz} = \times_{i=1}^n [(z_i-1)/N,z_i/N)$ for $\vz \in [N]^n$. For $\vx
= (x_1,\dots,x_n) \in \R^n$, let $\set{\vx} = \vx-\floor{\vx} \in [0,1)^n$ denote
the fractional part of $\vx$. Examine the sequence $\set{0\va},\set{1\va},\dots,\set{N^n
\va}$. Since the sequence has length $N^n+1$ and each element of the sequence lands
in one of the cubes in $\CC$, by the pigeonhole principle there must be distinct indices $l_1,l_2$, $0 \leq l_1 < l_2 \leq N^n$ and $\vz \in [N]^n$ such that
$\set{l_1\va},\set{l_2\va} \in I_\vz$. Since $I_\vz-I_\vz = (-1/N,1/N)^n$, we note that
$\length{\set{l_1\va}-\set{l_2\va}}_\infty < 1/N$. Let $l = l_2-l_1$ and $\va' =
\round{l\va}$, observing $1 \leq l \leq N^n$. For any $i \in [n]$, we have 
\[
|la_i-\round{la_i}| = \min_{k \in \Z} |la_i-k| \leq |(l_1-l_2)a_i-(\floor{l_1a_i}-\floor{l_2a_i})| = |\set{l_1a_i}-\set{l_2a_i}| < 1/N.
\]
In particular, $\length{l\va-\va'}_\infty = \length{l\va-\round{l\va}}_\infty <
1/N$, as needed. We now show that $\length{\va'}_\infty = l$. By
assumption on $\va$, there is a coordinate $i \in [n]$ such that $a_i = 1 =
\length{\va}_\infty$. Thus, $a'_i = \round{la_i} = l$ and $\length{\va'}_\infty
\geq l$. For any $j \in [n]$, also clearly have $l a_j \in [-l,l] \Rightarrow
a'_j = \round{la_j} \in [-l,l]$ since $l \in \N$. Thus, $\length{\va'}_\infty =
l$ as needed. 
\end{proof}

\begin{remark}
For $\va \in \R^n, \va' \in \Z^n, 1 \leq l \leq N^n$ as above, observe that $a_i = 0 \Rightarrow a'_i =
\round{la_i} = 0$. Furthermore, $\length{\va'}_\infty = l \leq N^n$. 
\end{remark}

\begin{definition}[Diophantine Approximation of Precision $N$]
For a vector $\va \in \R^n \setminus \set{0}$ and $N \geq 1$, we say that $\va'$
is  a precision $N$ Diophantine approximation of $\va$ if $\va'$ satisfies the
conditions of Lemma~\ref{lem:diophantine} on inputs $\va/\length{\va}_\infty$
and $N$. 

In the following, we will set $N = 10nR$, where $R$ is an integer upper bound on the
$\ell_1$ radius of the convex set $K \subseteq \R^n$ whose branching proof we
are modifying.

\end{definition}

\subsection{Farkas Certificates for General Convex Sets} 
\label{subsec:gen-farkas}

A Farkas certificate $\vLa \in \R^m_+$ certifies the infeasibility of the system $\MA \vx \leq \vb, \MA \in \R^{m \times n}, \vb \in \R^m$ if $\vLa^\Tr \MA = 0, \vLa ^\Tr \vb =-1$. It is possible to extend this definition to show a linear system is infeasible whenever $\vx \in K$ for a compact convex set $K$.

\begin{definition}[Generalized Farkas Certificate]\label{def:gen-Farkas}
Let $K \subseteq \R^n$ be a compact convex set, and $P := \set{\vx \in \R^n:
\MA \vx \leq \vb}, \MA \in \R^{m \times n}, \vb \in \R^m$. $\vLa \in
\R^m_+$ is a generalized Farkas certificate of infeasibility for $K \cap P$ if
$$\min_{\vx \in K} \vLa^\Tr (\MA \vx - \vb) > 0.$$
\end{definition}

\begin{lemma} With the notation of definition~\ref{def:gen-Farkas}, $K \cap P =
\emptyset$ if and only if there exists a generalized Farkas certificate $\vLa
\in \R^m_+$ of its infeasibility. Furthermore, if one generalized Farkas certificate exists, then so does one with at most $n+1$ non-zero coordinates.
\label{lem:gen-cert}
\end{lemma}

\begin{proof}
That a generalized Farkas certificate implies infeasibility is trivial.

Now let us suppose $K \cap P = \emptyset$. $K$ is compact and convex by
assumption, and $P$ is clearly closed and convex. Therefore, there exists a
strictly separating hyperplane $\vc \vx = d$ so that $K$ and $P$ lie on ``opposite sides''
of this hyperplane. More precisely, $\vc \vx - d > 0$ for $\vx \in K$, and $\vc
\vx - d < 0 $ for $\vx \in P$. 

$\vc \vx < d$ for every $\vx \in P$ means the system $\MA \vx \leq \vb, -\vc \vx
\leq -d$ is infeasible. Let $(\vLa, \gamma) \geq 0$ be a (conventional) Farkas
certificate of the infeasibility of this system: $\vLa ^\Tr \MA = \gamma \vc,
\vLa^\Tr \vb < \gamma d$. We now claim that $\vLa \geq 0$ is a generalized
Farkas certificate of infeasibility for $K \cap P$. Firstly, if $\gamma = 0$, we
have that $\min_{\vx \in K} \vLa^\Tr (\MA \vx - \vb) = -\vLa^\Tr
b > 0$. If $\gamma > 0$, then
\[
\vx \in K \Rightarrow \gamma(\vc \vx - d) > 0 \Rightarrow \vLa ^\Tr( \MA
\vx - \vb) > 0.
\]
In particular, $\min_{\vx \in K} \vLa^\Tr(\MA \vx - \vb) > 0$,
noting that the minimum is indeed achieved since $K$ is compact. 

By Caratheodory's theorem, there exists a generalized Farkas certificate of at most $n+1$ non-zero coordinates whenever a generalized Farkas certificate exists.

\end{proof}

Although the correctness of a conventional Farkas certificate can be verified
with simple matrix multiplication, this is not the case for a generalized Farkas
certificate. In particular, one must exactly solve the (convex) minimization problem in
definition~\ref{def:gen-Farkas} to verify the certificate. This is why the notion
of a certified branching proof is sensible only for specific classes of compact
convex sets, such as polyhedra.

The following lemma will be crucial for enabling us to deduce infeasibility
information for ``nearby'' polyhedra. The proof relies upon the existence of
generalized Farkas certificates as defined above.

\begin{lemma}
\label{lem:genFarkas}
Let $K \subseteq \R^n$ be a compact convex set and let $P = \set{\vx \in \R^n
: \MA \vx \leq \vb}, \MA \in \R^{m \times n}, \vb \in \R^m$, be a
polyhedron satisfying $P \cap K = \emptyset$. For $\vEps \in \R^m$, define
$P_{\vEps} := \set{\vx \in \R^n: \MA \vx \leq \vb + \vEps}$. Then, for any
$\vEps \in \R^m$, either $K \cap P_{\vEps} = \emptyset$, or there exists $j \in [m]$
such that $\eps_j > 0$ and $K \cap P_{\vEps-(n+1)\eps_j \ve_j} = \emptyset$.
\end{lemma}

\begin{proof}
We assume that $K \cap P_{\vEps} \neq \emptyset$, since otherwise there
is nothing to prove. 

Let $\vLa \in \R^m_+$ be a generalized Farkas certificate of infeasibility
for $K \cap P$ with at most $n+1$ non-zero coordinates as guaranteed by
Lemma~\ref{lem:gen-cert}. Let $j_* = \arg\max_{j \in [m]} \eps_j \lambda_j$. We
claim that $\eps_{j_*} \lambda_{j_*} > 0$.  Assume not, then $\eps_j \lambda_j \leq
0$ for all $i \in [m]$. In particular,
\begin{equation}
\min_{\vx \in K} \vLa^\Tr (\MA \vx - \vb - \vEps) = \min_{\vx \in K} \vLa^\Tr
(\MA \vx - \vb) - \vLa^\Tr \vEps  > - \vLa^\Tr \vEps \geq 0. \label{eq:gen-fark-1}
\end{equation}
Thus, $\vLa$ is a generalized Farkas certificate of infeasibility for $K \cap
P_\eps$. But this contradicts our assumption that $K \cap P_\eps \neq
\emptyset$. Therefore, we must have that $\eps_{j_*} \lambda_{j_*} > 0$. In
particular, since $\vLa \geq 0$, we have that $\eps_{j_*} > 0$ and $\lambda_{j_*} >
0$.  

We now show that $\vLa$ is in fact a valid generalized Farkas
certificate of infeasibility for $K \cap P_{\vEps-(n+1)\eps_{j_*} \ve_{j_*}}$.
Let $S = \set{j \in [m]: \lambda_j > 0}$, and note that by assumption $|S| \leq n+1$.
Using a similar calculation to~\eqref{eq:gen-fark-1}, we see that
\begin{multline*}
\min_{\vx \in K} \vLa^\Tr (\MA \vx - \vb - \vEps + (n+1)\eps_{j_*} \ve_{j_*}) > 
- \vLa^\Tr \vEps + (n+1)\eps_{j_*} \lambda_{j_*} \\ = - \sum_{j \in S} \eps_j
  \lambda_j +
(n+1) \eps_{j_*} \lambda_{j_*} \geq (n+1-|S|) \eps_{j_*} \lambda_{j_*} \geq 0.
\end{multline*}
Since $\vLa$ is a valid certificate of infeasibility, we have that $K \cap
P_{\vEps-(n+1) \eps_{j_*} \ve_{j_*}} = \emptyset$, as needed. 
\end{proof}

\subsection{Chv{\'a}tal-Gomory Cuts}

\begin{definition}[Chv{\'a}tal-Gomory Cut]
For $\va \in \Z^n$, the \emph{CG cut of $K$ induced by $\va$} is the
halfspace $H^{\rm cg}_K(\va) := H_{\va,\floor{h_K(\va)}}$. We define $\CG(K,\va)
:= K \cap H^{\rm cg}_K(\va)$ to be the result of applying the CG cut induced by
$\va$ to $K$. 

This definition is extended to an ordered list $\CL = (\va_1,\dots,\va_k)$ of
integer vectors as $\CG(K,\CL) := \CG(\CG(K,\va_1), (\va_2,\dots,\va_k))$.
That is, we first apply the CG cut induced by $\va_1$ to $K$ yielding
$\CG(K,\va_1)$, then we apply the CG cut induced by $\va_2$ to $\CG(K,\va_1)$
yielding $\CG(K,(\va_1,\va_2))$, and so forth. By convention, $\CG(K,\emptyset)
= K$, that is, applying the empty list of CG cuts does nothing to $K$.
\end{definition}

The following \emph{lifting lemma}, adapted from~\cite{DDV14}, shows that CG
cuts on a ``rational face'' $F$ of $K$ can be lifted to a CG cut of $K$ having
the same effect on the face. We note that lifting is also possible from
``irrational faces''~\cite{DDV14}, however this requires intersecting multiple
CG cuts to achieve the desired effect. The corresponding lemma for rational
polyhedra is classical~\cite{Chvatal73}. 

We include its proof for clarity and completeness. The proof follows the standard
approach of adding a large integer multiple of the normal vector to $F$ to the
cut.   

\begin{lemma}[Lifting CG cuts] 
\label{lem:cg-lift}
Let $K \subseteq \R^n$ be a non-empty compact set. Let $\vc \in \Z^n$, $F :=
F_K(\vc)$ and assume that $h_K(\vc) \in \Z$. Then for any $\va \in \Z^n$, there
exists $N \geq 0$ such that
\[
H^{\rm cg}_K(\va + i\vc) \cap H^=_K(\vc) =
H^{\rm cg}_F(\va) \cap H^=_K(\vc), \forall i \geq N.
\] 
\end{lemma}

For the proof, we will need the following technical lemma, which shows convergence
properties of a sequence of maximizing faces. 

\begin{lemma} 
\label{lem:max-lim-cpt}
Let $K \subseteq \R^n$ be a non-empty compact set. Let $(\va_i)_{i=1}^\infty
\in \R^n$ be a convergent sequence with $\va_\infty := \lim_{i \rightarrow
\infty} \va_i$ and let $F_i := F_K(\va_i)$, $i \in \N \cup \set{\infty}$. Then,
$\forall \eps > 0$ there exists $N_\eps \geq 1$ such that $\forall i \geq
N_\eps$, $F_i \subseteq F_\infty + \eps \B_1^n$.
\end{lemma}
\begin{proof}
For the sake of contradiction, let us assume that there exists a sequence
$(\vx_i)_{i=1}^\infty$ and an $\eps > 0$ such that $\vx_i \in F_i$
and $\vx_i \notin F_\infty + \eps \B_1^n$. Letting $K' = {\rm closure}(K
\setminus (F_\infty + \eps \B_1^n))$, we see that $K' \subseteq K$ is compact and
that $K' \cap F_\infty = \emptyset$. Furthermore, $\vx_i \in F_i \subseteq K'$,
$\forall i \in \N$. Therefore, by compactness of $K'$ there exists a convergent
subsequence $(\vx_{s_i})_{i=1}^\infty$ with limit point $\vy : = \lim_{i
\rightarrow \infty} \vx_{s_i} \in K'$. Note that by construction $\vy \in K$ and
$\vy \not\in F_\infty$. Since $K$ is compact, its support function $h_K$ is
continuous. By continuity of $h_K$ and the standard inner product, we conclude
that
\begin{align*}
\vv_\infty \vy &= \lim_{i \rightarrow \infty} \vv_{s_i} \vx_{s_i} 
               = \lim_{i \rightarrow \infty} h_K(\vv_{s_i}) \quad \left(\text{
since } \vx_{s_i} \in F_{s_i} \right) \\
               &= h_K(\vv_\infty).
\end{align*}
But then $\vy \in F_\infty$, a clear contradiction. The lemma thus follows.
\end{proof}

We now give the proof of the lifting lemma.

\begin{proof}[Proof of Lemma~\ref{lem:cg-lift}]
Firstly, if $\vc = 0$ then $F = K$ and the statement trivially holds for $N=0$.
Thus, we may assume that $\vc \neq 0$. 

Let $b = h_F(\va)$ and recall that $H^{\rm cg}_F(\va) = \set{\vx \in \R^n: \va
\vx \leq \floor{b}}$. For $i \geq 0$, let $b_i := h_K(\va + N\vc)
- i h_K(\vc)$. From here, we see that
\begin{align*}
\vx \in H^{\rm cg}_K(\va + i\vc) \cap H^=_K(\vc) &\Leftrightarrow 
(\va + i\vc) \vx = \floor{h_K(\va + i\vc) \vx}, \vc \vx = h_K(\vc) \\
&\Leftrightarrow (\va + i\vc) \vx \leq \floor{b_i + ih_K(\vc)}, \vc \vx = h_K(\vc) \\
&\Leftrightarrow (\va + i\vc) \vx \leq \floor{b_i}+ih_K(\vc), \vc \vx = h_K(\vc) \\
&\quad \quad \quad \quad \left(\text{since } ih_K(\vc) \in \Z \right) \\
&\Leftrightarrow \va \vx \leq \floor{b_i}, \vc \vx = h_K(\vc).
\end{align*}
Given the above, it suffices to show that there exists $N \geq 0$ such that
$\floor{b_i} = \floor{b}$, $\forall i \geq N$. Since $F$ is the set of
maximizers of $\vc$ in $K$, note that 
\[
b_i = h_K(\va + i \vc) - ih_K(\vc) \geq h_F(\va + i\vc) - ih_K(\vc) = h_F(\va) =
b, \forall i \geq 0.
\]
Letting $\eps_1 = \floor{b+1}-b > 0$, note that $\floor{b'} = \floor{b}$ for
$b' \in [b, b+\eps_1)$. Given this, it now suffices to show the
existence of $N \geq 0$ such that $b_i < b+\eps_1$, for $i \geq N$. Let $F_i :=
F_K(\va + i \vc)$, for $i \in N$. Since
$\va/i + \vc \rightarrow \vc$ as $i \rightarrow \infty$ and $K$ is compact, by
Lemma~\ref{lem:max-lim-cpt} for $\eps_2 > 0$ there exists $N_{\eps_2} \geq 0$
such that $F_i \subseteq F + \eps_2 \B_1^n$, for $i \geq N_{\eps_2}$. For $i \geq
N_{\eps_2}$, we may thus choose $\vx_i \in F_i$ and $\vy_i \in F$ satisfying
$\|\vx_i-\vy_i\|_1 \leq \eps_2$. From here, for $i \geq N_{\eps_2}$ we have
that
\begin{align*}
b_i &= h_K(\va + i \vc) - ih_K(\vc) = (\va + i \vc) \vx_i - ih_K(\vc) \quad
\left(\text{ since } \vx_i \in F_i \right) \\
    &\leq \va \vx_i + i h_K(\vc) - ih_K(\vc) = \va (\vx_i-\vy_i) + \va \vy_i
\quad
\left(\text{ since } \vx_i \in K \right)  \\
    &\leq \|\vx-\vy\|_1 \|\va\|_\infty + h_F(\va) \leq \eps_2 \|\va\|_\infty +
b \quad \left( \text{ since } \vy_i \in F \right).
\end{align*}
Setting $\eps_2 := \eps_1/(2\|\va\|_\infty)$ and
$N := N_{\eps_2}$ yields the desired bound. The lemma thus follows.
\end{proof}

\section{Bounding the coefficients of Branching Proofs}
\label{sec:coefficients}
In this section, we show how to transform any branching proof $\CT$ for a
compact convex set $K \subseteq R\B_1^n$ into a branching proof $\CT'$ having
small coefficients with length $\len{\CT'} = O(n\len{\CT})$.  
 
The construction of $\CT'$ is a two step process. In the first step, we
substitute each integer disjunction given by $(\va, b)$ by an approximation
$(\va', b')$ with coefficients of size $(nR)^{O(n^2)}$. This bounds
$\inner{\CT'}$ while keeping $\len{\CT'} = \len{\CT}$. We shall use the
``iterated Diophantine approximation'' technique introduced by Frank and
Tardos~\cite{FT87} to construct $\va', b'$ from $\va, b$.

It is possible that the new inequalities are ``stronger''; e.g., it is possible
that for $\va' \vx \leq b' \rimplies \va \vx \leq b$ and $\va' \vx \geq b'+1
\rimplies \va \vx \geq b+1$. However, one cannot always ensure this, and in
general we will only be able to guarantee that $\va' \vx \leq b' \rimplies \va
\vx \leq b + \eps$ and $\va' \vx \geq b'+1 \rimplies \va \vx \geq b+1-\eps$ for
some ``small'' $\eps > 0$. As explained in the introduction, the combined error
from all the substitutions may render the continuous relaxations at the leaves
nonempty. In a second step, we ``fix-up'' these newly feasible leaf nodes by
adding $O(n)$ judiciously chosen CG cuts to arrive at infeasible sets, causing
the $O(n)$ factor increase in $\len{\CT'}$. These cuts will be derived from
so-called valid substitution sequences (see Definition~\ref{def:validDioph}) of
the original disjunctions in $\CT$, which we construct together with the replacement
disjunctions $\va' \vx \leq b'$ or $\geq b'+1$ described above.

From here until the end of subsection~\ref{subsec:step1}, we explain the first
step, showing how to construct appropriate replacement disjunctions together
with substitution sequences and how to compute the initial (partial) replacement
tree $\CT'$ from $\CT$. In subsection~\ref{subsec:step2}, we explain the second
step, showing how to construct the requisite $O(n)$-size CP proof of
infeasibility for each leaf node of $\CT'$. Finally, in
subsection~\ref{subsec:step3}, we give the proof of Theorem~\ref{thm:coef-bnd}
which combines both steps.  

We begin with the following lemma, which collects the properties of Diophantine
approximations we will need to construct the replacement disjunctions and
substitution sequences. 

\begin{lemma}
\label{lem:approx dioph}
For any vector $\va  \in \R^n \setminus \set{0}, b \in \R, R, N \in \N$, let $\va'$ be
a Diophantine approximation of $\va $ of precision $N$, and let $\alpha =
\frac{\length{\va}_\infty}{\length{\va' }_\infty}$. Then the following statements hold:
\begin{enumerate}

\item [(i)] $\forall b' \in \R, \va '\vx  \leq b' \rimplies \va \vx  \leq \alpha \left(b' + \frac{R}{N}\right)$ and symmetrically, $\va '\vx  \geq b' \rimplies \va \vx  \geq \alpha \left( b' - \frac{R}{N} \right)$.

\item [(ii)] When $\frac{R}{N} < \frac{1}{4}, \alpha \geq 2$, we can uniquely
set $b' \in \Z$ according to exactly one of following cases:
\begin{itemize}

\item (non-$R$-dominating case): 
$-R\length{\va}_\infty-1 < b < R\length{\va}_\infty$ and $\exists$ unique $b' \in \Z, \len{b'} \leq R \lengthfit{\va'}_\infty,$ 
$$\text{ such that } (b, b+1) \cap \left[ \alpha \left(b'- \frac{R}{N} \right), \alpha \left(b' + \frac{R}{N} \right) \right] \neq \emptyset.$$

\item ($R$-dominating case): $\exists$ unique $b' \in \Z, -R \lengthfit{\va'}_\infty \leq b' \leq R \lengthfit{\va'}_\infty-1$,
$$\text{ such that } (b,b+1) \subseteq \left( \alpha
\left(b'+ \frac{R}{N} \right), \alpha \left(b' + 1 - \frac{R}{N} \right) \right),$$
or
$$b \geq R\lengthfit{\va}_\infty,  b' = R\lengthfit{\va'}_\infty,$$
or
$$b+1 \leq -R\lengthfit{\va}_\infty, b' = -R\lengthfit{\va'}_\infty-1.$$
\end{itemize}

Furthermore, in the $R$-dominating case we have that 
$$\va'\vx \leq b' \rimplies \va \vx \leq b \text{ and }  \va'\vx \geq b'+1 \rimplies \va \vx \geq b+1.$$

\end{enumerate}
\end{lemma}

\begin{proof}

\begin{enumerate}

\item [(i)] By definition of $\va '$, $\lengthfit{\frac{ \va}{\alpha}  - \va '}_\infty<1/N$. We have for any $b'\in \Z$:
$$\lengthfit{\vx }_1\leq R, \va '\vx  \leq b' \Rightarrow \frac{\va}{\alpha}
\vx  \leq b' + (\frac{ \va}{\alpha}  - \va ')\vx  \leq b' + \lengthfit{\frac{
\va}{\alpha}  - \va '}_\infty\lengthfit{\vx }_1 \leq b' + \frac{R}{N}.$$
Summarizing, we have that 
$$\lengthfit{\vx }_1 \leq R, \va '\vx  \leq b' \Rightarrow \va \vx  \leq \alpha \left( b' + \frac{R}{N}\right).$$

By a symmetric argument, we also have
$$\lengthfit{\vx }_1\leq R, \va '\vx  \geq b' \Rightarrow \va \vx  \geq \alpha \left( b' - \frac{R}{N} \right).$$

\item [(ii)] When $\frac{R}{N} < \frac{1}{4}$, the intervals of the form $I(b') := [\alpha
\left(b' - \frac{R}{N}\right), \alpha\left(b' + \frac{R}{N}\right)]$, $b' \in
\Z$, are pairwise disjoint. In fact, when $\alpha \geq 2$, they are more than unit distance apart. This implies that the interval $(b, b+1)$ cannot
intersect more than one of the intervals $I(b')$, $b' \in \Z$. 

Let us now suppose $-R\length{\va}_\infty-1 < b < R\length{\va}_\infty$. We now show that only $b' \in [-R \lengthfit{\va}_\infty, R\lengthfit{\va}_\infty] \cap \Z$ need be
considered in this case.

For $b' = -R \lengthfit{\va'}_\infty$, we have $I(b') = [R \alpha \left( \lengthfit{\va'}_\infty - \frac{1}{N} \right), R \alpha \left(
\lengthfit{\va'}_\infty + \frac{1}{N} \right)]$. The left end point
$-R\lengthfit{\va}_\infty - \frac{R \alpha}{N}$ of
$I(-R\lengthfit{\va'}_\infty)$ lies to the left of $b+1$ on the real line because $$- R
\lengthfit{\va}_\infty - \frac{R \alpha}{N} < -R\lengthfit{\va}_\infty <
b+1.$$ Similarly the right end point $R \lengthfit{\va}_\infty + \frac{R
\alpha}{N}$ of $I(R\lengthfit{\va'}_\infty)$ lies to the right of $b$
as $$R \lengthfit{\va}_\infty + \frac{R \alpha}{N} >
R\lengthfit{\va}_\infty > b.$$ Thus, either $(b, b+1)$ intersects some
$I_{b'}$ for $b' \in [-R\lengthfit{\va'}_\infty, R\lengthfit{\va'}_\infty] \cap \Z$ or
it lies in between two such consecutive intervals $I_{b'}, I_{b'+1}$: these are the
non-dominating and dominating cases respectively.

In the dominating case for $b \in (-R \lengthfit{\va}_\infty - 1, R\lengthfit{\va}_\infty)$, the fact that
$$(b,b+1) \subseteq \left( \alpha
\left(b'+ \frac{R}{N} \right), \alpha \left(b' + 1 - \frac{R}{N} \right) \right)$$
implies $b \geq \alpha \left(b' + \frac{R}{N}\right)$. Applying part (i) we see that 
$$\va' \vx \leq b' \rimplies \va \vx \leq \alpha \left(b' + \frac{R}{N}\right) \Rightarrow \va \vx \leq b.$$
On the other side, $b+1 \leq \alpha \left( b' + 1 - \frac{R}{N} \right)$ gives 
$$\va' \vx \geq b'+1 \rimplies \va \vx \geq \alpha \left(b' + 1 -
\frac{R}{N}\right) \Rightarrow \va \vx \geq b+1.$$

Now let us consider the situation where $b \geq R \lengthfit{\va}_\infty$. Then
$\forall \vx \in R \B_1^n$ we have $\va \vx \leq \lengthfit{\va}_\infty
\lengthfit{\vx}_1 \leq \lengthfit{\va}_\infty R \leq b$. Since this inequality
holds for every vector in $R \B_1^n$, we have that $\va'\vx \leq b' \rimplies
\va \vx \leq b$ $\forall b' \in \R$ and in particular for $b' = R
\lengthfit{\va'}_\infty$. Furthermore, for $b'= R \lengthfit{\va'}_\infty$,
$\va' \vx \geq b'+ 1$ does not hold for any $\vx \in R\B_1^n$ and thus $\va' \vx
\geq b'+1 \rimplies \va \vx\geq b+1$.

The symmetric reasoning applies to case $b+1\leq -R\lengthfit{\va}_\infty$. Setting
$b' = -R\lengthfit{\va}_\infty-1$, we firstly have that $\va \vx \geq b+1$ is a
valid inequality for $R\B_1^n$ and hence $\va' \vx \geq b'+1 \rimplies \va \vx
\geq b+1$ trivially. Secondly, the system $\va' \vx \leq b', \|\vx\|_1 \leq R$
is empty and hence $\va' \vx \leq b' \rimplies \va \vx \leq b$ trivially
as well.
\end{enumerate}
\end{proof}

In the sequel, we will say that $(\va',b')$ $R$-dominates or $R$-non-dominates
$(\va,b)$ when the corresponding case holds in Lemma~\ref{lem:approx dioph}. We
will also drop label $R$- when $R$ is clear from context. 

Part (ii) of the lemma above will be used to choose the right hand sides
$b_1,\dots,b_k$ given vectors $\va_1,\dots,\va_k$ that induce the pair of
inequality sequences (we think of one as the ``flipped'' version of the other)
$\va_1 \vx \leq b_1,\dots,\va_k \vx \leq b_k$ and $\va_1 \vx \geq
b_1,\dots,\va_{k-1} \vx \geq b_{k-1}, \va_k \vx \geq b_k+1$ respectively used to
approximate the left side $\va \vx \leq b$ and right side $\va \vx \geq b+1$ of
an initial disjunction. 


When applying the above lemma to an initial disjunction $\va \vx \leq b$ or
$\geq b+1$, the $R$-dominating case above is a scenario in which the naive
replacement of the disjunction $\va \vx \leq  b$ or $\geq b+1 \rightarrow
\va'\vx \leq b'$ or $\geq b'+1$ by the ``first pass'' Diophantine approximation
does the job. Indeed, if every branching decision in $\CT$ was dominated by its
first pass Diophantine approximation, then the naive replacements would be
sufficient to obtain a branching proof with bounded coefficients. 

Since domination does not always occur at the first level of approximation for
an original disjunction $\va \vx \leq b$ or $\geq b+1$, we will require the use
of an substitution sequence $\va_1 \vx \leq b_1, \dots, \va_1 \vx \leq b_k$ as
described previously. The exact properties needed from this sequence as well as the
algorithm to compute it are provided in the next subsection. At a high level, we
continue creating additional levels of approximation until the dominating case
occurs. More precisely, for every level $l \in [k]$, we will reduce the
inequality $\va \vx \leq b$ by subtracting non-negative combinations of the
equalities $\va_i \vx = b_i$, $i \in [l-1]$, to get a ``remainder inequality''
$\hat{\va}_l \vx \leq \hat{b}_l$. The remainder is then given together with its
precision $N$ Diophantine $\va_l$ as input to Lemma~\ref{lem:approx dioph} to
get the next level approximator $\va_l \vx \leq b_l$. The final iteration $k$
will correspond to the first time where the dominating case occurs (i.e., all
previous iterations are non-dominating).     

\begin{remark}
\label{rem:approx dioph}
Since we are interested in approximating not just an inequality but a
disjunction, it will be crucial that (non-)domination is well-behaved with
respect to both sides of the disjunction. For this purpose, we will heavily make
use of the following ``flip-symmetry'' in the definition of the $R$-domination
and $R$-non-domination. Namely, if $(\va', b')$ dominates $(\va, b)$, then
$(-\va', -b'-1)$ dominates $(-\va,-b-1)$, and if $(\va',b')$ non-dominates
$(\va,b)$ then $(-\va',-b')$ non-dominates $(-\va,-b-1)$. One can easily check
that these symmetries follow directly from simple manipulations of the
definitions. These symmetries are what will allow us to conclude that the
substitution sequence $\va_1 \vx \leq b_1, \dots, \va_k \vx \leq b_k$ and its
flipped version $\va_1 \vx \geq b_1, \dots, \va_k \vx \geq b_k+1$ will yield
good approximations to $\va \vx \leq b$ and $\va \vx \geq b+1$ respectively. 
\end{remark}

\subsection{Step 1: Replacing Large Coefficient Branches by Small Coefficient Approximations}
\label{subsec:step1}

Given a branching proof $\CT$ of infeasibility for $K \subseteq RB_1^n$, where
$R \in \N$, we begin the construction of the replacement proof $\CT'$ as
follows:  

We let $\CT'$ be a tree with vertex set $V'$ containing a vertex $v'$ for each
$v \in V[\CT]$, and an edge $e' = (v', w')$ for each $e = (v, w) \in E[\CT]$.
For each internal node $v \in V[T]$ with children $v_l,v_r$, we compute through
Algorithm~\ref{algo:step1} (see below) an approximation $(\va_{v'}, b_{v'})$ of
the disjunction $(\va_v, b_v)$ at $v$ of precision $R, N := 10nR, M := (10nR)^{n+2}$. We label the left
edge $e_l = (v',v'_l)$ in $\CT'$ by $\va_{v'} \vx \leq b_{v'}$ and the right
edge $(v',v'_r)$ by $\va_{v'} \vx \leq -b_{v'}-1$ (equivalently, $\va_{v'} \vx
\geq b'+1$). 

In this first phase of construction, note that $\CT'$ retains the same tree
structure as $\CT$. Furthermore, note that the output of Algorithm~\ref{algo:step1} must
serve equally well to approximate $\va_v \vx \leq b_v$ and $\va_v \vx \geq b_v
+1$ for $v \in \CT$.

The required properties of the replacements of the form $\va \vx \leq b
\rightarrow \va'\vx \leq b'$ are collected in the definition of a valid
substitution sequence defined below. A valid substitution sequence of $(\va, b)$
of precision $R, N, M$ consists of, along with the approximations $\va', b'$,
auxiliary information in the form of integers $k, b_1, b_2 \ldots b_k$, integer
vectors $\va_1, \ldots \va_k$ and nonnegative reals $\gamma_1, \ldots \gamma_k$.
While this auxiliary information is not included in the labels of $\CT'$, it is
computed by Algorithm~\ref{algo:step1} (and hence its existence is guaranteed).
Furthermore, the existence of the valid substitution sequence is crucial for
second part of the tree construction (see subsection~\ref{subsec:step2}), where
the inequalities from the substitution sequences are used to construct CP proofs
of infeasibility for the (possibly non-empty) leaves of $\CT'$ above. 

\begin{definition}[Valid Substitution Sequence]
\label{def:validDioph}
We define a \emph{valid substitution sequence} of an integer inequality $\va \vx \leq b$,
$\va \in \Z^n \setminus \set{0}$, $b \in \Z$ of precision $R, N, M  \in \N$ to be a list 
$$(\va', b', k, \va_1, b_1, \gamma_1,\dots, \va_k, b_k,
\gamma_k := 0),$$
where $k \in [n+1]$ and $\va', \va_i \in \Z^n, b', b_i \in \Z$,
$\gamma_i \in \R_+$, for $i \in [k]$, satisfying: 

\begin{enumerate}

\item $\lengthfit{\va'}_\infty \leq N^nM^{n+1}, \len{b'} \leq RN^nM^{n+1}$ and

$\lengthfit{\va_i}_\infty \leq 11nN^n, |b_i| \leq
R\lengthfit{\va_i}_\infty+1, i \in [k].$

\item For $l \in [k-1]$, we have
$$\va '\vx  \leq b',
\va _i \vx  = b_i, \forall~ i \in [l-1] \rimplies \va _l\vx  < b_l + 1.$$

\item For $l \in [k]$, we have
$$\va' \vx  \leq b',
\va _i \vx  = b_i, \forall~ i \in [l-1] \rimplies \va \vx  \leq b + \gamma_l.$$

\item For $l \in [k-1]$, we have
$$\va_{l} \vx  \leq b_{l}-1,
\va _i \vx  = b_i, \forall~ i \in [l-1] \rimplies \va \vx  \leq b - n \gamma_l.$$

\end{enumerate}
\end{definition}

\begin{remark}
In light of property $3$, the terms $\gamma_i$ are measures of precision for our
approximation of $\va \vx \leq b$. If $l_1 < l_2$, property $3$ when applied to
$l_2$ assumes more statements than when applied $l_1$. Intuitively, this suggests that
the implications should also be stronger (or at least not weaker). That is, 
one would expect 
$\gamma_{l_2} \geq \gamma_{l_1}$. Indeed, this assumption can be made without
loss of generality. More precisely, if
$(a',b',k,\va_1,b_1,\gamma_1,\dots,\va_k,b_k,\gamma_k := 0)$ is a valid
substitution, then so is
$(a',b',k,\va_1,b_1,\bar{\gamma}_1,\dots,\va_k,b_k,\bar{\gamma}_k)$, where
$\bar{\gamma}_i := \min_{j \in [i]} \gamma_i$.
\end{remark}

\LinesNumbered    
\begin{algorithm}
	\SetKwFunction{LongToShort}{LongToShort}
    \KwIn{%
         $\va \in \Z^n, b \in \Z, R, N, M \in \N$ such that $\frac{R}{N}<\frac{1}{4}$.
         }%

     \KwOut{%
				$\va' \in \Z^n , b' \in \Z, k \in \N$, and $\va_i \in \Z^n, b_i \in
\Z,\gamma_i \in \R_+$, $i \in [k]$, satisfying:
\begin{enumerate}
\item $(\va',b',k,\va_1,b_1,\gamma_1,\dots,\va_k,b_k,\gamma_k)$ is a valid
substitution sequence \newline of $\va \vx \leq b$ of precision $R,M,N$.
\item
$(-\va',-b'-1,k,-\va_1,-b_1,\gamma_1,\dots,-\va_{k-1},-b_{k-1},\gamma_{k-1},-\va_k,-b_k-1,\gamma_k)$
is a \newline valid substitution sequence of $-\va \vx \leq -b-1$ of precision $R,M,N$.
\end{enumerate}
        }%

         {\bf initialize} $\hat{\va}_1 = \va, \hat{b}_1 = b, j=1$;

    \While{$\lengthfit{\hat{\va}_j}_\infty > 10nN^n $}{%

        Set $\va _j$ as a Diophantine approximation of $\hat{\va }_j$ of precision $N$; 
        
        Apply Lemma~\ref{lem:approx dioph} part (ii) to $\hat{\va }_j, \hat{b}_j, \va_j, R, N$ to obtain $b_j$;
        
        \If{$(\va_j, b_j)$ dominates $(\hat{\va}_j, \hat{b}_j$)}
        {Set $k= j, \gamma_k = 0, \va ' = \sum_{i=1}^k M^{k -i} \va _i$ and $b' = \sum_{i=1}^k M^{k-i} b_i$;
        
        \KwRet{$\va', b', k, \va_i, b_i, \gamma_i, i \in [k]$}; \nllabel{alg1:ret1}
        }
        
        Set $\alpha_j = \frac{\lengthfit{\hat{\va }_j}_\infty}{\lengthfit{\va _j}_\infty}, \gamma_j = \frac{2\alpha_j}{5n}$;
        
        Set $\hat{\va }_{j+1} = \hat{\va }_j - \alpha_j\va _j, \hat{b}_{j+1} = \hat{b}_j - \alpha_jb_j$;
        
        Increment $j$;
     }
     Set $k=j, \gamma_k=0, \va_k = \va  - \sum_{i=1}^{k-1} \round{\alpha_i}\va _i, \tilde{b}_k = b - \sum_{i=1}^{k-1} \round{\alpha_i}b_i$;

     Set $b_k = \begin{cases} \tilde{b}_k&:\quad -R\lengthfit{\va_k}_\infty-1 < \tilde{b}_k < R\lengthfit{\va_k}_\infty \\
                          -R\lengthfit{\va_k}_\infty-1&:\quad \tilde{b}_k \leq -R\lengthfit{\va_k}_\infty-1 \\
                          R\lengthfit{\va_k}_\infty&:\quad R\lengthfit{\va_k}_\infty
\leq \tilde{b}_k \end{cases}$; \nllabel{alg1:b-check}

     Set $\va ' = \sum_{i=1}^k M^{k -i} \va _i$ and $b' = \sum_{i=1}^k M^{k-i} b_i$;
     
     \KwRet{$\va', b', k, \va_i, b_i, \gamma_i, i \in [k]$}; \nllabel{alg1:ret2}
     \caption{LongToShort($\va, b, R, N, M$)}
     \label{algo:step1}
     \end{algorithm}

\begin{lemma}
\label{lem:approx-props}
Algorithm~\ref{algo:step1} with input $\va, b, R, N, M$ such that
$N = 10nR, M = (10nR)^{n+2}$ terminates within $k\leq n+1$
iterations and outputs valid substitution sequences of $\va \vx \leq b$ and
$-\va \vx \leq -b-1$ of precision $R, N, M$.
\end{lemma}

\begin{proof}
We begin by showing that the number of coordinates of $\hat{\va }_{k+1}$ that
are zero is strictly greater than that of $\hat{\va }_k$. At iteration $j$, let
$p$ be such that $|(\hat{\va }_j)_p| = \lengthfit{\hat{\va }_j}_\infty$. As $\va
_j$ is the Diophantine approximation of $\hat{\va }_j$, we know that
$|(\frac{\hat{\va }_j}{\alpha_j} - \va _j)_p| = |(\frac{\lengthfit{\va
_j}_\infty}{\lengthfit{\hat{\va }_j}_\infty}\hat{\va }_j)_p - (\va_j)_p| <
1/10nR$. By assumption, $(\frac{\hat{\va }_j}{\lengthfit{\hat{\va
}_j}_\infty})_p = \pm 1$. Thus $(\frac{\lengthfit{\va
_j}_\infty}{\lengthfit{\hat{\va }_j}_\infty} \hat{\va }_j)_p \in \Z$, and so
$(\frac{\lengthfit{\va _j}_\infty}{\lengthfit{\hat{\va }_j}_\infty} \hat{\va
}_j)_p = (\va _j)_p$. As a result, $(\hat{\va}_{j+1})_p = 0$. As observed in remark 1.2, any zero entry of $\hat{\va }_j$ is also zero for $\hat{\va }_{j+1}$. 

By this reasoning, either Algorithm~\ref{algo:step1} terminates with $k \leq n$, or $\hat{\va}_{n+1}= 0$. The while loop terminates as $\lengthfit{\hat{\va}_{n+1}}_\infty \leq 10nN^n$. This proves that Algorithm~\ref{algo:step1} terminates within $n+1$ iterations.

We now show that Algorithm~\ref{algo:step1} outputs valid substitution
sequences. For this purpose, we first prove below that
$(\va',b',k,\va_1,b_1,\gamma_1,\dots,\va_k,b_k,\gamma_k)$ satisfies properties
1-4 of a valid substitution sequence of $\va \vx \leq b$ of precision $R,M,N$. After this, we
will argue that the flipped version of this sequence yields a valid substitution of
$-\va \vx \leq -b-1$ using the symmetries the algorithm. 

\begin{enumerate}

\item When Algorithm~\ref{algo:step1} returns from line~\eqref{alg1:ret1}, we
have $\lengthfit{\va_i}_\infty \leq N^n$, $i \in [k]$, since every $\va_i$ is
the result of Diophantine approximation of precision $N$. Furthermore, since
every $b_i$, $i \in [k]$, is then the output of Lemma~\ref{lem:approx dioph}
part (ii), we also have  $\len{b_i} \leq R\lengthfit{\va_i}_\infty+1 \leq
RN^n+1$,$i \in [k]$. Therefore, $$\lengthfit{\va'}_\infty \leq \sum_{i=1}^{k}
M^{k-i} \lengthfit{\va_i}_\infty \leq N^n \sum_{i=1}^{k} M^{k-i} \leq N^n
\sum_{i=0}^{n} M^{i} = N^n \frac{M^{n+1}-1}{M-1} \leq N^nM^{n+1}.$$

Similarly for $b'$, using $R,N \geq 1$ and $M \geq 3$, 
$$\lenfit{b'} \leq \sum_{i=1}^{k} M^{k-i} \lenfit{b_i} \leq (RN^n+1)
\sum_{i=1}^{k} M^{k-i} = (RN^n+1)\frac{M^{n+1}-1}{M-1} \leq RN^n M^{n+1}.$$

When Algorithm~\ref{algo:step1} returns from line~\eqref{alg1:ret2},
$\lengthfit{\va_i}_\infty \leq N^n$ for every $i \in [k-1]$ and
$\length{\hat{\va}_k} \leq 10N^n$. As in the previous case, we also have
$\len{b_i} \leq R\lengthfit{\va_i}_\infty+1$, $i \in [k]$. Note that for $i=k$,
this is enforced on line~\eqref{alg1:b-check} of the algorithm. Furthermore,
$b_k$ is indeed an integer since $R \in \N$ and $\tilde{b}_k,\lengthfit{\va_k}_\infty \in
\Z$ by construction. To bound
$\lengthfit{\va_k}_\infty$, we first note that
$$\lengthfit{\va_{k} - \hat{\va}_{k}}_\infty =
\lengthfit{\sum_{i=1}^{k-1} (\alpha_i - \round{\alpha_i})\va_i}_\infty
\leq \sum_{i=1}^{k-1} \lengthfit{(\alpha_i -
\round{\alpha_i})\va_i}_\infty \leq (k-1)N^n \leq  nN^n.$$
Since $\lengthfit{\hat{\va}_{k}}_\infty \leq 10nN^n$, we get that
$$\lengthfit{\va_{k}}_\infty \leq \lengthfit{\hat{\va}_{k}}_\infty + \lengthfit{\va_{k} - \hat{\va}_{k}}_\infty
\leq 10nN^n + nN^n = 11nN^n.$$
To bound $\length{\va'}_\infty$, by the triangle inequality
\begin{align}
\label{eq:a-prime-nb}
\lengthfit{\va'}_\infty &\leq \sum_{i=1}^{k-1} M^{k-i}\length{\va_i}_\infty + \length{\va}_k 
\leq N^n(\sum_{i=1}^n M^i + 11n) \\ &= N^n(\frac{M^{n+1}-1}{M-1}+11n-1) \leq
N^n(\frac{2M^{n+1}}{M-1}+11n-1) \leq N^nM^{n+1}, \nonumber
\end{align}
where it can be easily be checked that the last inequality holds for $M = (10nR)^{n+2}$ and $R,n
\geq 1$. The bound on $\len{b'}$ is computed in a manner similar to~\eqref{eq:a-prime-nb}: 

$$\len{b'} \leq \sum_{i=1}^k M^{k-i}(R\lengthfit{\va_i}_\infty+1) \leq RN^n(\frac{2M^{n+1}}{M-1}+11n-1) \leq RN^nM^{n+1}.$$

\item Let $l \in [k-1]$. When $\va _i \vx  = b_i, \forall~i \in [l-1]$, we have that
$$\va' \vx \leq b' \Leftrightarrow \sum_{i = l}^k M^{k-i}\va_i \vx \leq \sum_{i = l}^k M^{k-i}b_i \Leftrightarrow \va_{l} \vx + \sum_{i = l+1}^k M^{l-i}\va_i \vx \leq b_{l} + \sum_{i = l+1}^k M^{l-i}b_i.$$

By the proof of part 1, $\lengthfit{\va_i}_\infty \leq N^n, b_i \leq
R\lengthfit{\va_i}_\infty +1$, for $i \in [k-1]$. Using these bounds, we get that 
\begin{equation}
\label{eq:alg1-p2-1}
\va_{l}\vx + \sum_{i = l+1}^k M^{l-i}\va_i \leq b_{l} + \sum_{i = l+1}^k M^{l-i}b_i \rimplies \va_{l}\vx \leq b_{l} + \sum_{i = l+1}^k M^{l-i}b_i + R\lengthfit{\sum_{i = l+1}^k M^{l-i}\va_i}_\infty.
\end{equation}
The error in the last term is bounded by 
\begin{equation}
\label{eq:alg1-p2-2}
\sum_{i = l+1}^k M^{l-i}b_i + R\lengthfit{\sum_{i = l+1}^k M^{l-i}\va_i}_\infty
\leq (2R+1)N^n \left(\sum_{i=1}^{k-l} M^{-i} \right) \leq \frac{(2R+1)N^n}{M-1}
\leq \frac{1}{10n},
\end{equation}
where the last inequality is easily checked for $M = (10nR)^{n+2} = N^{n+2}$ and $n,R \in \N$.
Combining~\eqref{eq:alg1-p2-1} and~\eqref{eq:alg1-p2-2}, we conclude that
\begin{equation}
\label{eq:p2-final}
\va' \vx \leq b', \va _i \vx  = b_i, i \in [l-1] \rimplies \va_{l} \vx \leq b_{l} + \frac{1}{10n} < b_{l} + 1,
\end{equation}
as needed.

\item We first deal with the case $l=k$. We have $\va_k +
\sum_{i=1}^{k-1} M^{k-i}\va_i = \va'$ and $b_k +
\sum_{i=1}^{k-1} M^{k-i}b_i = b'$. So if $\va_i \vx = b_i$, $\forall~i \in
[k-1]$, we have that $\va' \vx \leq b' \Leftrightarrow \va_k \vx \leq b_k.$

When Algorithm~\ref{algo:step1} returns from line~\eqref{alg1:ret2}, we have
$\va_k + \sum_{i=1}^{k-1} \round{\alpha_i}\va_i = \va$ and $\tilde{b}_k +
\sum_{i=1}^{k-1} \round{\alpha_i}b_i = b$ by construction. By the same argument
as above, under the assumption $\va_i \vx = b_i, \forall~ i \in [k-1]$, we 
have that $\va_{k} \vx  \leq \tilde{b}_{k} \Leftrightarrow \va \vx \leq b$ .
It thus suffices to show that $\va_k \vx \leq b_k \rimplies \va_k \vx
\leq \tilde{b}_k$, recalling that $\gamma_k = 0$. This proceeds in an analogous fashion to the analysis of the
dominating case in Lemma~\ref{lem:approx dioph} part (ii). Firstly, by our
choice of $b_k$ on line~\eqref{alg1:b-check}, if
$-R\lengthfit{\va_k}_\infty-1 < \tilde{b}_k < R\lengthfit{\va_k}_\infty$ then $b_k = \tilde{b}_k$, so
this case is trivial. If $\tilde{b}_k \geq R\lengthfit{\va_k}_\infty$, then $b_k
= R\lengthfit{\va_k}$ and $\va_k \vx \leq \tilde{b}_k$ is valid inequality for
$R\B_1^n$. Thus, the implication $\va_k \vx \leq b_k \rimplies \va_k \vx
\leq \tilde{b}_k$ is again trivial. Lastly, if $\tilde{b}_k \leq
-R\lengthfit{\va_k}_\infty-1$, then $b_k = -R\lengthfit{\va_k}_\infty-1$ and the
system $\va_k \vx \leq b_k, \lengthfit{\vx}_\infty \leq R$ is empty. In
particular, $\va_k \vx \leq b_k \rimplies \va_k \vx \leq \tilde{b}_k$, as
needed.

Next, when the Algorithm~\ref{algo:step1} returns from line~\eqref{alg1:ret1},
by the guarantees of the $R$-dominating case in Lemma~\ref{lem:approx dioph} part
(ii), we have that
$$\va_k\vx \leq b_k \rimplies \hat{\va}_k \vx  \leq \hat{b}_k.$$
Similarly to the previous case, under the assumption $\va_i \vx = b_i$,
$\forall~i \in [k-1]$, we have that $\va' \vx \leq b' \Leftrightarrow \va_k \vx
\leq b_k$ and $\va \vx \leq b \Leftrightarrow \hat{\va}_k \vx \leq \hat{b}_k$.
The desired implication, $\va' \vx \leq b'$, $\va_i \vx = b_i$, $\forall~i \in
[k-1] \rimplies \va \vx \leq b$, thus follows.

Now suppose $l \in [k-1]$. From~\eqref{eq:p2-final} in part 2, we have that
$$\va' \vx \leq b', \va _i \vx  = b_i, i \in [l-1] \rimplies \va_{l} \vx \leq
b_l + \frac{1}{10n}.$$ 
By lemma~\ref{lem:approx dioph} part (i) applied to $\hat{\va}_{l+1}$ and $\va_{l+1}$,
\begin{equation}
\label{eq:alg-1-p3-1}
\va_{l} \vx \leq b_{l} + \frac{1}{10n} \rimplies \hat{\va}_{l}\vx \leq
\alpha_{l}\left(b_{l} + \frac{1}{10n} + \frac{R}{N}\right).
\end{equation}
Since $l \in [k-1]$, $\va_l, b_l$ non-dominates $\hat{\va}_l, \hat{b}_l$ and
thus by Lemma~\ref{lem:approx dioph} part (ii),
$$(\hat{b}_{l}, \hat{b}_{l} + 1) \cap \left[\alpha_{l}\left(b_{l}
- \frac{R}{N}\right), \alpha_{l}\left(b_{l}
+ \frac{R}{N}\right) \right] \neq \emptyset.$$
In particular, we get that
\begin{equation}
\label{eq:alg-1-p3-2}
\alpha_{l}\left(b_{l} + \frac{R}{N}\right) = \alpha_{l}\left(b_{l} -
\frac{R}{N}\right) + \alpha_l(\frac{2R}{N}) \leq \hat{b}_{l} + 1 + \alpha_l(\frac{2}{10n}). 
\end{equation}
Using $\alpha_l = \frac{\length{\hat{\va}_l}_\infty}{\length{\va_l}_\infty} \geq \frac{10 n N^n}{N^n}
= 10n$ combined with~\eqref{eq:alg-1-p3-2}, we get that
\begin{align}
\label{eq:alg-1-p3-3}
\alpha_{l}\left(b_{l}+\frac{1}{10n}+\frac{R}{n}\right) 
&\leq \hat{b}_l + 1 + \alpha_l \left(\frac{3}{10n} \right) 
\leq \hat{b}_l + \alpha_l \left(\frac{1}{10n}+\frac{3}{10n} \right) \\
&= \hat{b}_l + \alpha_l(\frac{2}{5n}) = \hat{b}_l + \gamma_l, \nonumber
\end{align}
where the last equality follows by definition of $\gamma_l$. Finally, under the
assumption that $\va_{i} \vx = b_i, \forall i \in [l-1]$, observe that for any 
$\delta \in \R$, we have that
\begin{equation}
\label{eq:alg-1-p3-4}
\hat{\va}_{l} \vx \leq \hat{b}_{l} + \delta \Leftrightarrow \va \vx \leq b + \delta.
\end{equation}

The desired implication $\va'\vx \leq b', \va_i \vx = b_i, \forall i \in [l-1]
\rimplies \va \vx \leq b + \gamma_l$ now follows directly from the above
combined with~\eqref{eq:p2-final},\eqref{eq:alg-1-p3-1} and~\eqref{eq:alg-1-p3-3}.

\item Repeating the argument from part $3$ starting from~\eqref{eq:alg-1-p3-1} with $\va_{l} \vx
\leq b_{l}-1$, we have that
\begin{equation}
\label{eq:alg-1-p4-1}
\va_{l} \vx \leq b_{l}-1 \rimplies \hat{\va}_{l} \vx \leq
\alpha_{l}(b_l-1+\frac{R}{N}).
\end{equation}
Using~\eqref{eq:alg-1-p3-3} in part 3, we see that
\begin{equation}
\label{eq:alg-1-p4-2}
\alpha_{l}(b_l-1+\frac{R}{N}) \leq \hat{b}_{l} - \alpha_l +
\alpha_l(\frac{2}{5n}-\frac{1}{10n}) = \hat{b}_l - \alpha_l(1-\frac{3}{10n}).
\end{equation}
Recalling $\gamma_l :=  \alpha_l(\frac{2}{5n})$, observe that $n \gamma_l =
\alpha_l(\frac{2}{5}) \leq \alpha_l(1-\frac{3}{10n})$ since $n \geq 1$. Combining together
with~\eqref{eq:alg-1-p4-1},~\eqref{eq:alg-1-p4-2} and~\eqref{eq:alg-1-p3-4} from
part 3, we conclude that
$$
\va\vx \leq b_l-1, \va_i \vx = b_i, \forall i \in [l-1] \rimplies \va \vx \leq b -
\alpha_l(1-\frac{3}{10n}) \leq b - n\gamma_l,
$$ 
as needed.
\end{enumerate}

We conclude the proof by showing that 
$$(-\va',-b'-1,k,-\va_1,-b_1,\gamma_1,\dots,-\va_{k-1},-b_{k-1},\gamma_{k-1},-\va_k,-b_k-1,\gamma_k)$$
is a valid substitution sequence of $-\va \vx \leq -b-1$ of precision $R,M,N$.
We have already proved that Algorithm~\ref{algo:step1} correctly outputs a valid
substitution sequence of $\va \vx \leq b$, so we are done if we show that $(-\va',-b'-1,k,-\va_1,-b_1,\gamma_1,\dots,-\va_{k-1},-b_{k-1},\gamma_{k-1},-\va_k,-b_k-1,\gamma_k)$ could have been output by Algorithm~\ref{algo:step1} upon input $-\va, -b-1$.

If $\va_i$ is a Diophantine approximation of $\hat{\va}_i$, then $-\va_i$ is a
Diophantine approximation of $-\hat{\va}_i$. Referring to Remark~\ref{rem:approx dioph} as our next step: when $j < k$,  $(\va_j, b_j)$ non-dominates $(\hat{\va}_j, \hat{b}_j)$, $(-\va_j, -b_j)$ also non-dominates $(-\hat{\va}_j, -\hat{b}_j-1)$. As a ratio of norms, the $\alpha_j$ values are identical for both executions of Algorithm~\ref{algo:step1}.

If Algorithm~\ref{algo:step1} with input $\va, b$ returned from line~\eqref{alg1:ret1}, then
the algorithm with input $-\va, -b-1$ must also return from line~\eqref{alg1:ret1}. This is
also a consequence of Remark~\ref{rem:approx dioph}: if $(\va_k, b_k)$ dominates
$(\hat{\va}_k, \hat{b}_k)$, then $(-\va_k, -b_k - 1)$ dominates $(-\hat{\va}_k,
-\hat{b}_k -1)$ and the algorithm returns with the expected valid substitution
sequence of $-\va, -b-1$.

If Algorithm~\ref{algo:step1} with input $\va, b$ returned from
line~\eqref{alg1:ret2}, this means $\lengthfit{\hat{\va}_k}_\infty \leq 10nN^n$.
Running Algorithm~\ref{algo:step1} with input $-\va, -b-1$ would give
$-\hat{\va}_k$, also obviously of small norm. Line $11$ would consequently give
$-\va - \sum_{i=1}^{k-1} \round{\alpha_i} (-\va_i) = -\va_k$ and $-b-1 -
\sum_{i=1}^{k-1} \round{\alpha_i}(-b_i) = -\tilde{b}_k-1$. It now suffices to
check that output of line~\eqref{alg1:b-check} given $-\tilde{b}_k-1$ is
$-b_k-1$, recalling that $b_k$ is the output of line~\eqref{alg1:b-check} on
input $\tilde{b}_k$. This follows by direct inspection, noting that it is
analogous to the ``flip-symmetry'' of Lemma~\ref{lem:approx dioph} part (ii).  
\end{proof}

Replacing branches with large coefficients with their valid approximations
reduces their bit-size, since the valid approximations have bit-size $O(n^3
\log_2
(2nR))$. However, we are not yet done. We do not yet have a valid branching
proof as the convex sets $K_{v'}$ associated to leaf nodes $v'$ of $\CT'$ are
not necessarily empty. We deal with this in Step 2.

\subsection{Step 2: Adding Chv{\'a}tal-Gomory (CG) Cuts to Trim the Leaves}
\label{subsec:step2}
We now show how to add CG cuts at each leaf of the current replacement tree
$\CT'$ for $\CT$, whose construction is described in the previous subsection, to
ensure that all the leaf nodes in the final tree have empty continuous
relaxations. The final tree will simply simulate the effect of the CG cuts
applied to the leaves of $\CT'$ using additional branching decisions (see the
proof of Theorem~\ref{thm:coef-bnd} in the next subsection).  

Recall from the last subsection, that every leaf node $v \in \CT$ has an associated
leaf $v' \in \CT'$ in the current replacement tree.  The continuous relaxation
for $v'$ is $K_{v'} = P_{v'} \cap K$ (recall that unlike $K_v :=
P_v \cap K$, $K_{v'}$ need not be empty), where the inequalities defining $P_{v'}$ are derived from
\emph{valid substitution sequences} (as in Definition~\ref{def:validDioph}) of the
original defining inequalities for $P_v$. Given this setup, our task is to add
``low-weight'' CG cuts to $P_{v'} \cap K$ to derive the empty set. 

The main result of this subsection is a general procedure for deriving such CG cuts
for any polyhedron $P'$ induced by valid substitution sequences of the defining
inequalities of a polyhedron $P$, where $P$ satisfies $K \cap P = \emptyset$.
The procedure will return a list of at most $2(n+1)$ CG cuts, which is
responsible for the $O(n)$ factor blowup in the final tree size. Second, the
normals of these CG cuts will all come from the substitution lists for the
inequalities defining $P$, which ensures that they have low weight. The formal
statement of this result is given below:   


\begin{theorem}
\label{thm:step2}
Let $K \subseteq R \B_1^n$, $R \in \N$, be a compact convex set, and let $P =$ 
$\set{\vx \in \R^n: \MA \vx \leq \vb }$, $\MA \in \Z^{m \times n}$, $\vb \in
\Z^m$, be a polyhedron satisfying $P \cap K = \emptyset$. For each defining
inequality $\va_i \vx \leq b_i$ of $P$, for $i \in [m]$, let $(\va_i', b_i',
k_i, \va_{i, 1}, b_{i,1}, \gamma_{i,1}, \dots, \va_{i, k_i}, b_{i, k_i},
\gamma_{i,k_i})$ be a valid substitution sequence of precision $R, N := 10nR, M
:= (10nR)^{n+2}$. Let $P' = \set{\vx \in \R^n: \MA' \vx \leq \vb'}$ be the
corresponding ``substitution'' polyhedron, where $\MA' \in \Z^{m \times n}$ has
rows $\va'_1,\dots,\va'_m$ and $\vb' \in \Z^m$ has rows $b'_1,\dots,b'_m$. 

Then, there exists an ordered list $\CL := (\va_{j_1,p_1},-\va_{j_1,p_1},\dots, \va_{j_l,p_l},-\va_{j_l,p_l}) \subseteq
\Z^n$, where $j_r \in [m]$ and $p_r \in [k_{j_r}-1]$, $r \in [l]$, satisfying
$\CG(K \cap P', \CL) = \emptyset$ and $|\CL| = 2l
\leq 2(n+1)$.
\end{theorem}

\begin{proof}
To prove theorem~\ref{thm:step2}, we give a procedure to construct such a list
$\CL$ in Algorithm~\ref{algo:step2}. To prove the theorem, it thus suffices to
prove the correctness of Algorithm~\ref{algo:step2}.

\LinesNumbered
\begin{algorithm}
    \SetKwFunction{GenCGCuts}{GenCGCuts}

    \KwIn{%
 $K, P := \MA \vx \leq \vb, P' := \MA' \vx \le \vb', R,M,N \in \N$, $(\va_i',
b_i', k_i, \va_{i, 1}, b_{i,1},\gamma_{i,1},\dots, \va_{i, k_i}, b_{i, k_i},
\gamma_{i,k_i})$, for $i \in [m]$, a valid substitution sequence of $\va_i\vx \leq b_i$  of
precision $R,M,N$, as in theorem ~\ref{thm:step2}.
		 }%

    \KwOut{%
        An ordered list $\CL :=
(\va_{j_1,p_1},-\va_{j_1,p_1},\dots,\va_{j_l,p_l},-\va_{j_l,p_l})$ satisfying \newline
$\CG(K \cap P', \CL) = \emptyset$ and $0 \leq l \leq n+1$.
        }%

         {\bf initialize} $\CL =\emptyset$, $V = \R^n$, $p(i) =1$, for $i \in
[m]$, and $\vEps = (\gamma_{1,p(1)},\dots,\gamma_{m,p(m)})$;
         
	Define $P_{\vEps}:= \set{\vx \in \R^n: \MA \vx \leq \vb + \vEps}$;

   \While{$K \cap P_{\vEps} \neq \emptyset$ and $V \neq \emptyset$} 
    {\nllabel{alg2:while} Apply Lemma~\ref{lem:genFarkas} to $K, P, \vEps$ to obtain $j_* \in [m]$ satisfying
$\eps_{j_*} > 0$ and $K \cap P_{\vEps-(n+1)\eps_{j_*} \ve_{j_*}} = \emptyset$; \nllabel{alg2:j}
    
  Append vectors $\va_{j_*, p(j_*)},-\va_{j_*, p(j_*)}$ to the list $\CL$;

	Update $V \leftarrow V \cap \set{\vx \in \R^n: \va_{j_*, p(j_*)} \vx = b_{j_*, p(j_*)}}$; \nllabel{alg2:V}

	\For{$j$ from $1$ to $m$} {Increment $p(j)$ to the largest integer $p \in
[k_j]$ satisfying $V \subseteq \set{\vx \in \R^n: \va_{j, i} \vx = b_{j, i}, 1 \leq i < p}$; \nllabel{alg2:maximal} 

  $\eps_j \leftarrow \gamma_{j,p(j)}$;\nllabel{alg2:eps} 
	}
}
     \KwRet{$\CL$};

     \caption{Generate CG Cuts}
     \label{algo:step2}
     \end{algorithm}

To begin, we first give a high level description of the algorithm and explain
the key invariants it maintains. The algorithm proceeds in
iterations, associated with runs of the while loop on line~\ref{alg2:while}. At
each iteration, we append the pair of CG cuts induced by $\va_{j,p},-\va_{j,p}$, $j
\in [m]$, $p \in [k_j-1]$, from one of our substitution lists to the end
$\CL$. These cuts are chosen so that after adding them to $\CL$, we can
guarantee that $\CG(K \cap P', \CL)$ satisfies the equality $\va_{j,p} \vx =
b_{j,p}$. 

We keep track of these learned equalities using the affine subspace $V \subseteq
\R^n$, which is initialized as $V = \R^n$ at the beginning of the algorithm. The
principal invariant needed to prove correctness of the algorithm is as follows: 
at the beginning of an iteration $l \geq 1$, $\CL,V$ satisfy
\[
(i)\quad \CG(K \cap P', \CL) \subseteq V \quad \text{ and } \quad \dim(V) \leq n-l+1.
\]
The condition on the dimension of $V$ above will be achieved by ensuring that
the new equality we add is not already implied by $V$. Precisely, the dimension
of $V$ will decrease by at least one at every iteration where we pass the while
loop check. Using (i), at the beginning of iteration $l=n+2$ (i.e., after $n+1$
iterations) we will have that $\dim(V) \leq -1$ and hence $\CG(K \cap P',\CL)
\subseteq V = \emptyset$. In particular, the while loop check $V \neq \emptyset$
will fail and we will correctly terminate. Thus, assuming (i) holds, the algorithm always
terminates after at most $n+1$ iterations. Since we add only $2$ CG cuts per
iteration, the total number of cuts in the list $\CL$ will be at most $2(n+1)$ by
the end the algorithm. To prove that (i) holds, we first introduce two other
important invariants.  

To keep track of the learned equalities in each substitution list, we keep a
counter $p(j) \in [k_j]$, for $j \in [m]$. For the second invariant, the
algorithm maintains that at the beginning of each iteration we have that 
\[
(ii)\quad V = \cap_{j \in [m]} \cap_{1 \leq i \leq p(j)-1} \set{\vx \in \R^n: \va_{j,i} \vx = b_{j,i}},
\]
and that each $p(j) \in [k_j]$,$j \in [m]$, is maximal subject to the above
equality. That is, for each $j \in [m]$, $V$ satisfies all the equalities
$\va_{j,i} \vx = b_{j,i}$ for $i \in [p(j)-1]$, and, if $p(j) < k_j$, $V$ does
not satisfy $\va_{j,p(j)} \vx = b_{p(j)}$. The counters are initialized to
$p(1)=\dots=p(m)=1$ corresponding to $V = \R^n$ (i.e., we have not yet learned
any equalities), which indeed yields a maximal choice. That the affine space $V$
can expressed in the above form is a simple consequence of how we update it on
line~\eqref{alg2:V}. Namely, we only update $V$ when we add the equality $\va_{j_*,p(j_*)} \vx =
b_{j_*,p(j_*)}$ to $V$ on line~\eqref{alg2:V}. Note that since $\eps_{j_*} > 0$ on line~\eqref{alg2:j},
we must have $p(j_*) < k_{j_*}$, since otherwise $\eps_{j_*} =
\gamma_{j_*,k_{j_*}} = 0$ (by definition of a valid substitution sequence). Thus, we only
add an inequality $\va_{j,p} \vx = b_{j,p}$, $j \in [m]$, to $V$ if $1 \leq p < k_j$ and if
$V$ satisfies $\va_{j,i} \vx = b_{j,i}$ for all $i \in [p-1]$, as needed. Lastly,
the required maximality is directly ensured by line~\eqref{alg2:maximal}. This
proves that (ii) is indeed maintained. Note that under maximality, for each $j
\in [m]$ such that $p(j) < k_j$, adding the equality $\va_{j,p(j)} \vx =
b_{p(j)}$ to $V$ must reduce the dimension of $V$ by at least one (more
precisely, adding this equality either makes $V$ empty or reduces its dimension
by exactly $1$). We will use this in the proof of (i).

With this notation, we may state the final invariant, which will be a direct
consequence of the first two and the definition of a valid substitution sequence. Letting
$\vEps := (\gamma_{1,p(i)},\dots,\gamma_{m,p(m)})$ denote the ``error level''
for each constraint of $P$ (note that this equality is maintained on
line~\eqref{alg2:eps}), at the beginning of each iteration we maintain 
\[
(iii) \quad \CG(K \cap P',\CL) \subseteq P_{\vEps},
\]
where $P_{\vEps} := \set{\vx \in \R^n: \MA \vx \leq \vb+\vEps}$. Crucially,
invariant (iii) justifies the first termination condition $K \cap P_{\vEps} =
\emptyset$, since if this occurs $\CG(K \cap P',\CL) \subseteq K \cap P_{\vEps} =
\emptyset$. Note that for a constraint $j \in [m]$, with $p(j) = k_j$, the
effective error level $\eps_j = \gamma_{j,k_j} = 0$ (by definition of valid
substitution). That is, we have effectively ``learned'' the 
defining constraint $\va_j \vx \leq b_j$ for $P$ for any $j \in [m]$ with
$p(j)=k_j$. Clearly, once all the constraints of $P$ have been learned, we will
have $\CG(K \cap P',\CL) \subseteq K \cap P = \emptyset$, where the last equality
is by assumption. 

We now show that (iii) is a consequence of (i) and (ii). Let $\CL$,$V$,$p$ and
$\vEps$ be the state at the beginning of some iteration $l \geq 1$, and
assume that (i) and (ii) hold. Then, for each $j \in [m]$, we have that
\begin{align}
\CG(K \cap P',\CL) &\subseteq K \cap P' \cap V 
                     \quad \left(\text{ by (i) and } \CG(K \cap P',\CL) \subseteq K
\cap P'\right) \label{eq:inv-2} \\
&\subseteq R\B_1^n \cap \set{\vx \in \R^n: \va'_j \vx \leq b'_j,
\va_{j,i} \vx = b_{j,i}, \forall~ i \in [p(j)-1]} \nonumber \\
& \quad \quad \quad\quad \left(\text{ by (ii) and }K \subseteq R\B_1^n \right)  \nonumber \\
                 &\subseteq \set{\vx \in \R^n: \va_j \vx \leq b_j + \gamma_{j,p(j)}} 
                            \quad \left(\text{ by Definition
~\ref{def:validDioph} part 3. }\right). \nonumber
\end{align}
Since the above holds for all $j \in [m]$, and $\eps_j = \gamma_{j,p(j)}$, for $j
\in [m]$, this proves invariant (iii).

Given the above, to prove correctness of algorithm it suffices to establish invariant (i).
We now show that invariant (i) holds by induction on the iteration $l \geq 1$. Let
$\CL$,$V$,$p$ and $\vEps$ denote the state at the beginning of some iteration $l
\leq 1$ for which (i) holds. Note that (i) trivially holds for the base case
$l=1$ since $V = \R^n$. By the reasoning in the previous paragraphs, we also
have that invariant (ii) and (iii) hold at the beginning of $l$. We must now
show that (i) holds at the beginning of iteration $l+1$ under these assumptions.
Clearly, we may assume that we pass the while loop check $K \cap P_\eps
\neq \emptyset$ and $V \neq \emptyset$, since otherwise there is nothing to
prove.    

Let $j_* \in [m]$ be the index satisfying $\eps_{j_*} > 0$ and $K \cap P_{\vEps
- (n+1)\eps_{j_*} \ve_{j_*}} = \emptyset$ as guaranteed by
  Lemma~\ref{lem:genFarkas}. This index indeed exists since we already checked
that $K \cap P_{\vEps} \neq \emptyset$. As argued for (ii), we also know that
$p(j_*) < k_{j_*}$, which will ensure we have access to the required inequalities from
the valid substitution sequence of $\va_{j_*} \vx \leq b_{j_*}$. Letting $P'_\CL := \CG(K
\cap P',\CL)$, to prove that (i) holds for $l+1$,  it now suffices to show that
\begin{align*}
(a)&~ \CG(P'_\CL,(\va_{j_*,p(j_*)},-\va_{j_*,p(j_*)})) \subseteq \set{\vx \in \R^n: \va_{j_*,p(j_*)} \vx =  b_{j_*,p(j_*)} }, \\
(b)&~ \dim(V \cap \set{\vx \in \R^n: \va_{j_*,p(j_*)} \vx \leq b_{j_*,p(j_*)}}) \leq \dim(V)-1.
\end{align*}
As explained previously, (b) follows directly the maximality assumption in (ii)
and $p(j_*) < k_{j_*}$.
We may thus focus on (a). To begin, using (i) and (ii) and the same analysis as
in~\eqref{eq:inv-2}, we see that
\begin{align*}
P'_\CL &\subseteq R\B_1^n \cap \set{\vx \in \R^n: \va'_{j_*} \vx \leq b'_{j_*},
\va_{j_*,i} \vx = b_{j_*,i}, \forall i \in [p(j_*)-1]} \\
     &\subseteq \set{\vx \in \R^n: \va_{j_*,p(j_*)} \vx < b_{j_*,p(j_*)}+1},
\end{align*}
where the last containment follows from Definition~\eqref{def:validDioph} part 2.
In particular,
\begin{equation}
\label{eq:a-1-1}
\sup_{\vx \in P'_\CL} \va_{j_*,p(j_*)} \vx < b_{j_*,p(j_*)}+1 \Rightarrow
\floor{\sup_{\vx
\in P'_\CL} \va_{j_*,p(j_*)}\vx } \leq b_{j_*,p(j_*)},
\end{equation}
since $b_{j_*,p(j_*)} \in \Z$. From~\eqref{eq:a-1-1}, we conclude that 
\begin{equation}
\label{eq:a-1}
\CG(P'_\CL,\va_{j_*,p(j_*)}) \subseteq \set{\vx \in \R^n: \va_{j_*,p(j_*)} \vx
\leq b_{j_*,p(j_*)}}.
\end{equation}
From here, again using (i) and (ii), we have that
\begin{align}
P'_\CL \cap &\set{\vx \in \R^n: \va_{j_*,p(j_*)} \vx \leq b_{j_*,p(j_*)}-1} \nonumber \\
&\subseteq R\B_1^n \cap \set{\vx \in \R^n: \va_{j_*,p(j_*)} \vx \leq b_{j,p(j_*)}-1,
\va_{j_*,i} \vx = b_{j_*,i}, \forall~i \in [p(j_*)-1]} \nonumber \\
&\subseteq \set{\vx \in \R^n: \va_{j_*} \vx \leq b_{j_*}-n\eps_{j_*}}, \label{eq:a-2-1} 
\end{align}
where the last containment follows from Definition~\eqref{def:validDioph} part
4 and $\eps_{j_*} = \gamma_{j_*,p(j_*)}$. Noting that 
\[
P_{\vEps} \cap \set{\vx \in \R^n: \va_{j_*} \vx \leq b_{j_*}-n\eps_{j_*}} =
P_{\vEps-(n+1)\eps_{j_*}\ve_{j_*}},
\] 
by the guarantees of Lemma~\ref{lem:genFarkas}, invariant (iii)
and~\eqref{eq:a-2-1}, we therefore have that
\[
P'_\CL \cap \set{\vx \in \R^n: \va_{j_*,p(j_*)} \vx \leq b_{j_*,p(j_*)}-1} 
\subseteq K \cap P_{\vEps-(n+1)\eps_{j_*}\ve_{j_*}} = \emptyset.
\]
In particular, we must have that
\begin{equation}
\label{eq:a-2-2}
\sup_{\vx \in P'_\CL} -\va_{j_*,p(j_*)} \vx < -b_{j_*,p(j_*)}+1 \Rightarrow
\floor{\sup_{\vx \in P'_\CL} -\va_{j_*,p(j_*)} \vx} \leq -b_{j_*,p(j_*)},
\end{equation}
since $-b_{j_*,p(j_*)} \in \Z$. From~\eqref{eq:a-2-2}, we conclude that
\begin{equation}
\label{eq:a-2}
\CG(P'_\CL, -\va_{j_*,p(j_*)}) \subseteq \set{\vx \in \R^n: -\va_{j_*,p(j_*)} \vx \leq -b_{j_*,p(j_*)}}.
\end{equation}
Property (a) now follows directly by combining~\eqref{eq:a-1}
and~\eqref{eq:a-2}. This concludes the proof of invariant (i) and the proof of
correctness of the algorithm. 
\end{proof}

\subsection{Proof of Theorem~\ref{thm:coef-bnd}}
\label{subsec:step3}

Let $N = 10nR, M = (10nR)^{n+2}$. Given $\CT$, we construct $\CT'$ a labeled
binary tree with the same structure as that of $\CT$ as described at the
beginning of subsection~\ref{subsec:step1}. Recall that for each internal node
$v \in \CT$ with associated disjunction $\va_v \vx \leq b_v$ or $\geq b_v+1$, we retrieve a pair of
valid substitution sequences from $\LongToShort(\va_v, b_v, R, N, M)$, yielding
the precision $R,N,M$ sequence $(\va'_{v},
b'_{v},k_v,\va_{v,1},b_{v,1},\gamma_{v,1},\dots,$ $\va_{v,k},b_{v,k},\gamma_{v,k})$
for $\va_v \vx \leq b_v$ and the corresponding flip (as in the output description of
Algorithm~\ref{algo:step1}) for $-\va_v \vx \leq -b_v-1$. For the corresponding
node $v' \in \CT'$, we create two children $v'_l,v'_r$ and label the left edge
$(v',v'_l)$ with $\va'_v \vx \leq b'_v$ and the right edge $(v',v'_r)$ with
$-\va'_v \vx \leq -b'_v-1$.

From the properties of a valid substitution sequence, we have that
$\lengthfit{\va'_{v}}_\infty \leq N^nM^{n+1} $ and $\len{b'_{v}} \leq RN^n
M^{n+1}$. The choice of $N = 10nR, M = (10nR)^{n+2}$ gives
$$\lengthfit{\va'_{v}}_\infty \leq (10nR)^n (10nR)^{(n+2)(n+1)} = (10nR)^{n^2 + 4n
+2} \quad \text{ and } \quad \len{b'_{v}} \leq R (10nR)^{n^2 + 4n +2}.$$ Both of these quantities are
upper bounded by $(10nR)^{(n+2)^2}$. For $\vx \in \Z^n, \inner{\vx} \leq n + n
\log_2 (1+\lengthfit{\vx}_\infty)$, so the bit-size of each inequality is
$O(n^3\log_2(2nR))$. 

Consider an arbitrary leaf node $v' \in \CT'$ with associated leaf node $v \in
\CT$. Observe that by construction 
$$P_{v'} =
\set{\vx \in \R^n : \va'_e \vx \leq b'_e, e \in E[P_{\CT'}(v')]}$$
satisfies the hypotheses of Theorem~\ref{thm:step2}, so that there exists a
list $\CL_{v'}$ of integer vectors such that $\CG(K \cap P', \CL_{v'}) =
\emptyset$.  More precisely, $\CL_{v'} =
(\va_{j_1,p_1},-\va_{j_1,p_1},\dots,\va_{j_l,p_l},-\va_{j_l,p_l})$, where $l
\leq (n+1)$ and $\va_{j_r,p_r}, j_r \in [m_v], p_r \in [k_j-1]$, $r \in [l]$,
are taken from the valid substitution
sequences of precision $R,N,M$ of the inequalities in the system $\MA_v \vx \leq
\vb_v$, $\MA_v \in \Q^{m_v \times n}$, $\vb_v \in \R^{m_v \times n}$, defining
$P_v$. Note that by the properties of an $R,M,N$ valid substitution sequence,
$\lengthfit{\va_{j_r,p_r}}_\infty \leq 11nN^n, \len{b_{j_r,p_r}}\leq R11nN^n+1$,
$r \in [l]$, and hence via the same argument as above each
$(\va_{j,r},b_{j,r})$ can be described using $O(n^2 \log_2(2nR))$ bits.

We now explain how to extend $\CT'$ to a valid branching proof. For each leaf
node $v' \in \CT'$, we will build a branching proof of infeasibility for
$K_{v'}$ of length $O(n)$ which simulates the effect of the CG cuts in
$\CL_{v'}$. By appending these sub-branching proofs to $\CT'$ below each leaf
node $v'$, the extended $\CT'$ clearly becomes a valid branching proof for $K$
having length at most $|\CT'| = O(n) |\CT|$ by construction. 
 
The construction of the subtree at $v'$ using $\CL_{v'} =
(\va_{j_1,p_1},-\va_{j_1,p_1},\dots,\va_{j_l,p_l},-\va_{j_l,p_l})$ (as above)
proceeds as follows. Starting from $v'$, we create two children $v'_l,v'_r$, and
label the edge $(v',v'_l)$ with the inequality $\va_{j_1,p_1} \vx \leq
b_{j_1,p_1}$ and the edge $(v',v'_r)$ with the inequality $\va_{j_1,p_1} \vx
\geq b_{j_1,p_1}+1$. Recall that by the definition of a CG cut, the continuous
relaxation $K_{v'_r}$ at the right child $v'_r$ is now empty. The construction
now proceeds inductively on $v'_l$ using the sublist
$(-\va_{j_1,p_1},\dots,\va_{j_l,p_l},-\va_{j_l,p_l})$.  Note that for every cut
in $\CL'$, we add a left and right child to the current left-most leaf of the
partially constructed subtree, for which the continuous relaxation of the newly
added right child is always empty. At the end of the construction, it is easy to
see that the left-most leaf of the constructed subtree has $\CG(K_{v'}, \CL')$
as its continuous relaxation, which is empty by assumption. From here, we
immediately get that the constructed subtree yields a valid branching proof of
infeasibility for $K_{v'}$, and that the number of nodes in the subtree distinct
from $v'$ is exactly $2|\CL_{v'}| \leq 4(n+1)$. Furthermore, we may bound the
bit-size of this subtree by $O(n^3 \log_2(2nR))$, since it has $O(n)$ nodes and
every edge is labeled with an inequality of bit-size $O(n^2 \log_2(2nR))$. 

To bound the total bit-size $\inner{\CT'}$ of the final branching proof $\CT'$,
we combine the bit-size bound from the subtrees above together with the total
bit-size of all the replacement disjunctions of the form $a'_v \vx \leq b'_v$
or $\geq b'_v+1$ (as above) labeling the outgoing edges of nodes in $\CT'$
associated with internal nodes of $\CT$. Given that each disjunction $a'_v \leq
b'_v$ or $\geq b'_v+1$ requires $O(n^3 \log_2(2nR))$ bits as explained above,
their total bit-size is bounded by $O(n^3 \log_2(2nR) |\CT|)$. Furthermore, the
bit-size contribution from all the subtrees in $\CT'$ associated with leaf nodes
of $\CT$ is $O(n^3 \log_2(2nR) |\CT|)$, since the number of these subtrees is
bounded by $|\CT|$ and each has bit-size $O(n^3 \log_2(2nR))$ as explained above.
Thus, the total bit-size $\inner{\CT'} = O(n^3 \log_2(2nR) |\CT|)$
as needed. 

\subsection{Proof of Corollary~\ref{cor:coef-bnd-pol}}
The primary tools for this proof will be the following well-known facts
pertaining to the bit-size of linear programs: (see ~\cite{Schrijver86} Chapter 10 for a thorough treatment):

\begin{lemma}
\label{lem:LP-bitsizes}
Let $\MA \in \Q^{m \times n}, \vb \in \Q^m, \vc \in \Q^n$.
\begin{enumerate}

\item For $\vc \in \Q^n$, if $\max \set{\vc \vx : \MA \vx \leq \vb }$ is finite, then it has size at most $4(\inner{\MA, \vb} + \inner{\vc})$.

\item If the system $\vLa \MA = 0, \vLa \geq 0$ and $\vLa \vb < 0$ is
feasible, then there exists a solution $\vLa$ of with bit-size
$\inner{\lambda} = O(n \inner{\MA})$. 


\end{enumerate}
\end{lemma}

\begin{remark}
Part 2 of the above lemma in fact corresponds to a bound on the bit-size of a
generator $\vLa$ of the relevant extreme ray of the cone $\vLa \MA = 0, \vLa
\geq 0$, where we note that a Farkas certificate of infeasibility for $\MA \vx
\leq \vb$ (if it exists) can always be chosen to be an extreme ray of this cone.
This is also the reason why the bit-size bound does not in fact depend on $\inner{\vb}$. 
\end{remark}

\begin{proof}[Proof of Corollary~\ref{cor:coef-bnd-pol}]
Since $K = \set{\vx \in \R^n: \MC \vx \leq \vd}$ is a polytope (i.e., bounded),
we may invoke lemma~\ref{lem:LP-bitsizes} part 1 to conclude that $$\max_{\vx
\in P} \lengthfit{\vx}_1 = \max_{\vy \in \set{-1,1}^n} \max \set{ \vy \vx : \MC
\vx \leq \vd } \leq 2^{4(L + 2n)}$$ using that $\inner{\vy} \leq 2n$, for $\vy
\in \set{-1,1}^n$, and that $|a|\leq 2^{\inner{a}}~ \forall a \in \Q$. Therefore,
$K \subseteq R\B_1^n$ for $R=2^{4L + 8n}$. Theorem~\ref{thm:coef-bnd} applied
with $R = 2^{4L + 8n}$ already gives us a $\CT'$ with $|\CT'| \leq O(n |\CT|)$,
where every inequality $\va'_e \vx \leq b'_e$, for $e \in E[\CT']$, satisfies
$\inner{\va'_e, b'_e} \leq O(n^3\log_2(2nR))) = O(n^{3}L)$ and $\inner{\CT'} =
O(n^3 \log_2(2nR)) |\CT|) = O(n^3 L |\CT|)$. However, $\CT'$ is not yet a
\emph{certified} branching proof. 

We are done if we can add to each leaf node $v' \in \CT'$ a Farkas certificate
$\vLa_{v'}$ of small size. Recall the continuous relaxation at $v'$ in $\CT'$ is
$K_{v'} = \set{\vx \in \R^n: \MC \vx \leq \vd, \MA_{v'} \vx \leq \vb_{v'}}$. By
assumption, we know that $K_{v'} = \emptyset$, and thus by Farkas's Lemma there
exists $\vLa_{v'} := (\vLa_{v',1},\vLa_{v',2}) \geq 0$ such that $\vLa_{v',1}
\MC + \vLa_{v',2} \MA_{v'} = 0$ and $\vLa_{v',1} \vd + \vLa_{v',2} \vb_{v'} < 0$.
Thus, by lemma~\ref{lem:LP-bitsizes} part 2, there exists a solution $\vLa_{v'}$
whose bit-complexity is upper bounded by $O(n \inner{\MC,\MA_{v'}})$. This
quantity may be large since we have not controlled the number of rows
$\MA_{v'}$.  By Caratheodory's theorem however, there exists a solution
$\vLa_{v'}$ with at most $n+1$ non-zero entries. Therefore, we can restrict our
attention to a subset of the rows of $\MC$ and $\MA_{v'}$ of cardinality at most
$n+1$. As argued above, by Theorem~\ref{thm:coef-bnd} each row of $\MA_{v'}$ has
bit-size at most $O(n^3 \log_2(2nR)) = O(n^3 L)$, and by assumption $\inner{\MC}
\leq L$. Thus, by restricting to the appropriate sub-system, the bit-length of
the non-zero entries of $\vLa_{v'}$ can be bounded by $O(n((n+1)n^3 L + L)) = O(n^5
L)$, as needed. 

Since the number of nodes in $|\CT'| = O(n |\CT|)$, the combined bit-size of the
Farkas certificates above is at most $O(n^6 L |\CT|)$. This dominates the
contribution of the disjunctions to the bit-size of $\CT'$, which by
Theorem~\ref{thm:coef-bnd} is $O(n^3 L |\CT|)$. Thus, the certified version of
has size $\inner{\CT'} = O(n^6 L |\CT|)$, as needed.
\end{proof}


\section{Simulating Enumerative Branching Proofs by Cutting Planes}
\label{sec:simulation}

In this section, we prove that enumerative branching proofs can be simulated by
CP, and give an application to Tseitin formulas (see subsection~\ref{subsec:tseitin}).

To begin, we first extend the lifting lemma (Lemma~\ref{lem:cg-lift}) to
a sequence of CG cuts. This will allow us to use induction on subtrees of an
enumerative branching proof.   

\begin{lemma}[Lifting Sequences of CG cuts]
\label{lem:cg-seq-lift} 
Let $K \subseteq \R^n$ be a non-empty compact set. Let $\vc \in \Z^n$, $F :=
F_K(\vc)$ and assume that $h_K(\vc) \in \Z$. Let $\va_1,\dots,\va_k \in \Z^n$.
Then, there exists $n_1,\dots,n_k \geq 0$ such that
\[
\CG(K, (\va_1+n_1 \vc,\dots, \va_k + n_k \vc))
\cap H^=_K(\vc) = \CG(F,(\va_1,\dots,\va_k)) \text{ .} 
\]
\end{lemma}
\begin{proof}
We prove the statement by induction on $i$. For $i = 0$, there are no CG cuts to
apply and the statement becomes $K \cap H^=_K(\vc) = F$, which follows by
definition. For $i \in [k]$, we assume the induction hypothesis
\begin{equation}
\label{eq:cg-seq-1}
K_{i-1} \cap H^=_K(\vc) = F_{i-1},
\end{equation}
where 
\[
K_{i-1} := \CG(K, (\va_1+n_1\vc,\dots,\va_{i-1}+n_{i-1}\vc)) \quad \text{ and }
\quad F_{i-1} := \CG(F, (\va_1,\dots,\va_{i-1})).
\]
We must prove the existence of $n_i \geq 0$ such that~\eqref{eq:cg-seq-1} holds
for $i$. Firstly, if $F_{i-1} = \emptyset$, then regardless of the choice of
$n_i \geq 0$, both the sets $F_i$ and $K_i \cap H^=_K(\vc)$ will be empty since they are both contained in $F_{i-1} = \emptyset$.
In particular, we may set $n_i = 0$ and maintain the desired equality. 

So assume $F_{i-1} \neq \emptyset$. From here, since $\emptyset \neq F_{i-1}
\subseteq F$, where we recall that $F$ is the set maximizers of $\vc$ in $K$,
and $F_{i-1} \subseteq K_{i-1} \subseteq K$, we have that   
\[
h_K(\vc) \geq h_{K_{i-1}}(\vc) \geq h_{F_{i-1}}(\vc) = h_K(\vc) \in \Z.
\]
In particular, $H^=_{K_{i-1}}(\vc) = H^=_{K}(\vc)$. Therefore, by the induction
hypothesis~\eqref{eq:cg-seq-1}
\[
K_{i-1} \cap H^=_{K_{i-1}}(\vc) = K_{i-1} \cap
H^=_K(\vc) = F_{i-1} \text{ ,}
\]
that is to say, $F_{i-1}$ is the set of maximizers of $\vc$ in $K_{i-1}$. Furthermore, since
$K_{i-1}$ is the intersection of $K$ with closed halfspaces and $K$ is compact,
$K_{i-1}$ is also compact. Therefore, we may apply Lemma~\ref{lem:cg-lift} to
choose $n_i \geq 0$ satisfying
\begin{equation}
\label{eq:cg-seq-2}
H^{\rm cg}_{K_{i-1}}(\va_i + n_i \vc) \cap H^=_K(\vc) = 
H^{\rm cg}_{F_{i-1}}(\va_i) \cap H^=_K(\vc).
\end{equation}
We use the above $n_i$ to define 
\[
K_i := K_{i-1} \cap H^{\rm cg}_{K_{i-1}}(\va_i + n_i \vc)
= \CG(K,(\va_1+n_1 \vc,\dots,\va_i+n_i\vc)).
\]
Intersecting both sides of~\eqref{eq:cg-seq-2} with $K_{i-1}$, we conclude that
\begin{align*}
H^{\rm cg}_{K_{i-1}}(\va_i + n_i \vc) \cap K_{i-1} \cap H^=_K(\vc) &= H^{\rm
cg}_{F_{i-1}}(\va_i) \cap K_{i-1} \cap H^=_K(\vc) \\ 
\Leftrightarrow K_i \cap H^=_K(\vc) &= H^{\rm cg}_{F_{i-1}}(\va_i) \cap F_{i-1}
\\ 
\Leftrightarrow K_i \cap H^=_K(\vc) &= F_i,
\end{align*}
as needed. The lemma thus follows.
\end{proof}

We are now ready to prove the main result of this section, which shows that
enumerative branching proofs can be simulated by CP.  

\begin{proof}[Proof of Theorem~\ref{thm:branching-to-cp}]
Our procedure for converting enumerative branching proofs to CP proofs is given
by Algorithm~\ref{alg:branch-to-cp}. The proof of correctness of the procedure
will yield the theorem:

\begin{claim} 
Given an enumerative branching proof $\CT$ of integer infeasibility for a compact
convex set $K \subseteq \R^n$, Algorithm~\ref{alg:branch-to-cp} correctly
outputs a list $\CL = (\va_1,\dots,\va_N) \in \Z^n$ satisfying $\CG(K,\CL) =
\emptyset$ and $|\CL| := N \leq 2|\CT|-1$.
\end{claim}
\begin{proof}
~
\subparagraph*{Algorithm Outline} 
We first describe the algorithm at a high level and then continue with a formal
proof. The procedure traverses the tree $\CT$ in order, visiting the children of
each node from right to left. We explain the process starting from the root node
$r \in \CT$. To begin, we examine its branching direction $\va_r \in \Z^n$ and
bounds $l_r \leq u_r$ satisfying 
\[
\set{\va_r \vx: \vx \in K} \subseteq [l_r,u_r],
\]
recalling that $r$ has a child $r_b$ for each $b \in [l_r,u_r] \cap \Z$. 

Starting at $r$, the procedure adds CG cuts to ``chop off'' the children of $r$
moving from right to left. In particular, it alternates between adding the CG
induced by $\va_r$ to $K$, which will either make $K$ empty or push the
hyperplane $H^=_{\va_r,b}$, where $b:=h_K(\va_r)$, to the next child of $r$, and
recursively adding CG cuts induced by the subtree $\CT_{r_b}$ rooted at the
child $r_b$ of $r$. The cuts computed on the subtree $\CT_{r_b}$ will be used to
chop off the face $K \cap H^=_{\va_r,b}$ from $K$, which will require lifting
cuts from the face to $K$ using Lemma~\ref{lem:cg-seq-lift}. Once the face has
been removed, we add the CG cut induced by $\va_r$ to move to the next
child. The process continues until all the children have been removed and $K$ is
empty.   

\LinesNumbered

\begin{algorithm}
    \SetKwFunction{EnumToCP}{EnumToCP}
    
    \KwIn{Compact convex set $K \subseteq \R^n$, Enumerative branching proof $\CT$ for $K$.}
    \KwOut{List $\CL$ of CG cuts satisfying $\CG(K,\CL) = \emptyset$ and $|\CL| \leq
2|\CT|-1$.}

    {\bf initialize} $\CL = \emptyset$, $r \leftarrow $ root of $\CT$ \;

    \If{$K = \emptyset$} {
      \KwRet{$\emptyset$}\;
    }

    Retrieve branching direction $\va_r \in \Z^n$ and bounds $l_r \leq u_r$\; 
    $K \leftarrow \CG(K,\va_r)$\; \nllabel{bcp-init-cut}
    $\CL \leftarrow (\va_r)$\; 
    \While{$l_r \leq h_K(\va_r)$}{ \nllabel{bcp-while}
      $b \leftarrow h_K(\va_r)$\;
      $\CT_{r_b} \leftarrow $ subtree of $\CT$ rooted at $r_b$\; \nllabel{bcp-subtree}
      $N' \leftarrow \EnumToCP(F_K(\va_r), \CT_{r_b})$\; \nllabel{bcp-face-rec}
      $N \leftarrow$ Lift CG cuts in $N'$ from $F_K(\va_r)$ to $K$ using
Lemma~\ref{lem:cg-seq-lift}\; \nllabel{bcp-face-lift}
      $K \leftarrow \CG(K,N)$\; \nllabel{bcp-face-cut}
      $K \leftarrow \CG(K,\va_r)$\; \nllabel{bcp-v-cut}
      Append $N,\va_r$ to $\CL$\; \nllabel{bcp-cut-add} 
    }
    \KwRet{$\CL$}\;
    \caption{EnumToCP($K$, $\CT$)}%
     \label{alg:branch-to-cp}
     \end{algorithm}

\subparagraph*{Analysis} We show correctness by induction on $|\CT| \geq 1$.
Let $r \in \CT$ denote the root node with branching direction $\va_r \in \Z^n$
and bounds $l_r \leq u_r$.

We prove the base case $|\CT|=1$. If $K = \emptyset$, then no CG cuts are needed
and clearly $0 \leq 2|\CT|-1 = 1$. If $K \neq \emptyset$, then letting $r$ be
the root node, we must have $[l_r,u_r] \cap \Z = \emptyset \Rightarrow
\floor{u_r} < l_r$. Since $h_K(\va_r) \in [l_r,u_r]$, the initializing CG cut we
add on line~\ref{bcp-init-cut} induced by $\va_r$ will make $K$ empty. This
follows since after the cut $h_K(\va_r) \leq \floor{u_r} < l_r$. The algorithm
thus correctly returns $\CL = (\va_r)$, where $|\CL| = 1 = 2|\CT|-1$, as needed. 

Now assume that $|\CT| \geq 2$ and that the algorithm is correct for all
smaller trees. If $K = \emptyset$ or we do not enter the while loop on
line~\ref{bcp-while}, the algorithm correctly returns by the above analysis.
So we now assume that $K \neq \emptyset$ and that the algorithm performs at
least one iteration of the while loop. 

Let $K_0$ denote the state of $K$ at the beginning of the algorithm. Let $K_i$,
$b_i$, $\CL_i$, for $i \geq 1$, denote the state of $K$, $b$, $\CL$ at the beginning
of the $i^{\rm th}$ iteration of the while loop on line~\ref{bcp-while} and let
$N_i$, $i \geq 1$, denote the state of $N$ at the end of the $i^{\rm th}$ iteration.
Let $T \geq 1$ denote the last iteration (that passes the check of the
while loop). By the design of the algorithm, it is direct to check that $K_i =
\CG(K_0,\CL_i)$, $\CL_{i+1} = (\CL_i, N_i, \va_r)$ and that $b_i = h_{K_i}(\va_r)$,
$\forall i \in [T]$, where we define $\CL_{T+1}$ to be the list of CG cuts
returned by the algorithm, $K_{T+1} := \CG(K_0,\CL_{T+1}) = \emptyset$ and
$b_{T+1} := h_{K_{T+1}}(\va_r) = -\infty$. Note also that $K_i \neq \emptyset$,
$\forall i \in [T]$, since otherwise we would have terminated earlier.  

To begin, we claim that 
\begin{equation}
\label{eq:bcp-1}
b_i \in [l_r,u_r] \cap \Z, \forall i \in [T].
\end{equation}
To see this, note first that for any compact convex set $\CC$ either
$\CG(\CC,\va_r) = \emptyset$ and $h_{\CG(\CC,\va_r)}(\va_r) = -\infty$ or
$\CG(\CC,\va_r) \neq \emptyset$ and $h_{\CG(\CC,\va_r)}(\va_r) =
\floor{h_{\CC}(\va_r)} \in \Z$, where the latter claim follows from convexity
and compactness of $\CG(\CC,\va_r)$. Since we apply the CG cut induced by
$\va_r$ to $K$ directly before the while loop and at the end of every iteration,
we immediately get that $b_i = h_{K_i}(\va_r) \in \Z$, $i \in [T]$.
Furthermore, since $\emptyset \neq K_i \subseteq K_0$, $\forall i \in [T]$, and
$\set{\va_r \vx: \vx \in K_0} \subseteq [l_r,u_r]$, we must also have $b_i \in
[l_r,u_r]$.

Given~\eqref{eq:bcp-1}, for each $i \in [T]$, we see that $r_{b_i}$ is indeed a
child of $r$. Let $\CT_{r_{b_i}}, i \in [T]$ denote the subtree of $\CT$ rooted
at $r_{b_i}$. We  claim that
\begin{equation}
\label{eq:bcp-2}
b_{i+1} \leq b_i-1, i \in [T], \quad \text{ and } \quad |N_i| \leq
2|\CT_{r_{b_i}}|-1, i \in [T].
\end{equation}
To see this, first recall that $\CT_{r_{b_i}}$ is a branching proof for $K_0
\cap H^=_{\va_r,b_i}$. In particular, since $K_i \subseteq K_0$, $\CT_{r_{b_i}}$
is also a valid branching proof for $K_i \cap H^=_{\va_r,b_i} = F_{K_i}(\va_r)$.
By the induction hypothesis the call to 
\EnumToCP$(F_{K_i}(\va_r),\CT_{r_{b_i}})$
on line~\ref{bcp-face-rec} therefore correctly returns a list $N'_i$ of CG cuts
satisfying $\CG(F_{K_i}(\va_r),N'_i) = \emptyset$ and $|N'_i| \leq
2|\CT_{r_{b_i}}|-1$. Furthermore, by Lemma~\ref{lem:cg-seq-lift}, the lifting $N_i$
of $N'_i$ to $K$ computed on line~\ref{bcp-face-lift} satisfies $|N_i| = |N'_i|
\leq 2|\CT_{r_i}|-1$ and $\CG(K_i,N_i) \cap H^=_{\va_r,b_i} =
\CG(F_{K_i}(\va_r),N'_i) = \emptyset$, as needed. Letting $K'_i = \CG(K_i,N_i)$, by
compactness of $K'_i$ we therefore must have $h_{K'_i}(\va_r) < b_i$. Recalling
that $K_{i+1} = \CG(K'_i,\va_r)$, we see that $b_{i+1} = h_{K_{i+1}}(\va_r) \leq
\floor{h_{K'_i}(\va_r)} \leq b_i-1$, as needed.

From the above, we see that the procedure clearly terminates in finite time and
returns a list $\CL_{T+1}$ satisfying $\CG(K_0,\CL_{T+1}) = \emptyset$. It remains
to bound the size of $|\CL_{T+1}|$. Since we add $1$ CG cut before the while loop,
and at iteration $i \in [T]$, we add $|N_i|+1$ CG cuts, the total number of
cuts is
\[
1 + \sum_{i=1}^T (|N_i|+1) \leq 1 + \sum_{i=1}^T 2|\CT_{r_{b_i}}| \leq 2|\CT|-1,
\]
where the last inequality follows since the sum is over subtrees rooted at
distinct children of $r$, noting that $|\CT| = 1 + \sum_{b \in [l_r,u_r] \cap
\Z} |\CT_{r_b}|$. This completes the proof. 
\end{proof}
\end{proof}

\subsection{Upper Bounds for Tseitin Formulas} 
\label{subsec:tseitin}

\begin{proof}[Proof of Theorem~\ref{thm:tseitin}]
As explained in the introduction, given Theorem~\ref{thm:branching-to-cp}, it
suffices to show that the Beame et al~\cite{BFIKPPR18} SP refutation is in fact
enumerative. We thus describe their refutation briefly to make clear that this is
indeed the case.

We start with a Tseitin formula indexed by a graph $G = (V,E)$, of maximum degree
$\Delta$, together with parities $l_v \in \{0,1\}$, $v \in V$, satisfying
$\sum_{v \in V} l_v \equiv 1 \mod 2$. We recall that the variables $\vx \in
\{0,1\}^E$ index the corresponding subset of edges where the assignment $\vx$ is
a satisfying assignment iff $\sum_{e \in E: v \in e} x_e \equiv l_v \mod 2$,
$\forall v \in V$. 

The Beame et al refutation proceeds as follows. At the root node $r$, we first
divide the vertex set $V = V^r_1 \cup V^r_2$ arbitrarily into two parts of
near-equal size. We then branch on the number of edges crossing the cut 
\[
x(E[V^r_1,V^r_2]) := \sum_{\{v_1,v_2\} \in E, v_1 \in V^r_1, v_2 \in V^r_2} x_{{v_1,v_2}} \in
\set{0,\dots,|E[V^r_1,V^r_2]|}.
\]
Let $c$ be the child with $x(E[V^r_1,V^r_2]) = b$, for $b \in
\set{0,\dots,|E[V^r_1,V^r_2]|}$. $c$ chooses $i_c \in
\{1,2\}$ such that $\sum_{v \in V^r_{i_c}} l_v \neq b \mod 2$, corresponding
the set of vertices still containing a contradiction. From here, again $c$
partitions $V^r_{i_c} = V^c_1 \cup V^c_2$ into two near-equal pieces. 
We now branch twice: we first branch on the number of edges crossing the cut
$x(E[V^c_1,V^c_2])$, creating corresponding children, and at each such child, we
branch on number of edges crossing the cut $x(E[V^c_1,V \setminus V^c_1])$. From
here, every child $c'$, two levels down from $c$, can decide which set of
vertices $V^c_1$ or $V^c_2$ still contains a contradiction. The process
continues in a similar way until we find a contradicting set corresponding to a
single vertex $v$. At this point, one constructs a complete branching tree on
all possible values of the edges outgoing from $v$. This completes the
description.

It is clear from the description, that every branching decision is enumerative.
As shown in Beame et al, the above SP refutation has length $2^\Delta(n
\Delta)^{O(\log n)}$. Theorem~\ref{thm:branching-to-cp} shows that one can convert
it to a CP refutation of at most twice the length. This completes the proof. 
\end{proof}

\bibliographystyle{alpha}
\bibliography{branching}

\end{document}